\documentclass[utf8]{frontiersSCNS} % for Science, Engineering and Humanities and Social Sciences articles

\usepackage{soul}
\usepackage{amssymb}
\usepackage{booktabs, bbm}
\usepackage{multirow}
\usepackage[T1]{fontenc}% optional T1 font encoding
\usepackage{amsthm}
\usepackage{amsmath}
\usepackage[flushleft]{threeparttable}
\usepackage{graphicx,array,xcolor}
\usepackage{xspace, comment}
\usepackage{tikz}
\usepackage{xr} 
\usepackage{caption}
\usepackage{subcaption}

\usepackage{url,hyperref,lineno,microtype,subcaption}
\usepackage[onehalfspacing]{setspace}

\graphicspath{{./plots/}}

\newcommand{\etal}{\mbox{\emph{et al.}}\xspace}

\newcommand{\HONN}{\mbox{$\mathop{\mathtt{HONN}}\limits$}\xspace}
\newcommand{\SMM}{\mbox{$\mathop{\mathtt{SMM}}\limits$}\xspace}
\newcommand{\hkSVM}{\mbox{$\mathop{\mathtt{hkSVM}}\limits$}\xspace}

\newcommand{\NetMHCtrd}{\mbox{$\mathop{\mathtt{NetMHC3.0}}\limits$}\xspace}
\newcommand{\NetMHCfth}{\mbox{$\mathop{\mathtt{NetMHC4.0}}\limits$}\xspace}

\newcommand{\mhcflurrynd}{\mbox{$\mathop{\mathtt{MHCflurry 2.0}}\limits$}\xspace}

\newcommand{\hingeone}{\mbox{$\text{H}_v$}\xspace}
\newcommand{\hingetwo}{\mbox{$\text{H}_l$}\xspace}
\newcommand{\hingethree}{\mbox{$\text{H}_i$}\xspace}
\newcommand{\meansquare}{\mbox{MS}\xspace}

\newcommand{\convmodel}{\mbox{$\mathop{\mathtt{ConvM}}\limits$}\xspace}
\newcommand{\spannyconvmodel}{\mbox{$\mathop{\mathtt{SpConvM}}\limits$}\xspace}
\newcommand{\mhcflurry}{\mbox{$\mathop{\mathtt{MHCflurry}}\limits$}\xspace}

\newcommand{\avgrank}{\mbox{$\mathop{\mathtt{AR}}$}\xspace}
\newcommand{\avgrankone}{\mbox{$\mathop{\mathtt{AR}_{\scriptsize{100}}}$}\xspace}
\newcommand{\avgrankfive}{\mbox{$\mathop{\mathtt{AR}_{\scriptsize{500}}}$}\xspace}
\newcommand{\hr}{\mbox{$\mathop{\mathtt{HR}}$}\xspace}
\newcommand{\hrone}{\mbox{$\mathop{\mathtt{HR}_{\scriptsize{100}}}$}\xspace}
\newcommand{\hrfive}{\mbox{$\mathop{\mathtt{HR}_{\scriptsize{500}}}$}\xspace}
\newcommand{\auc}{\mbox{$\mathop{\mathtt{AUC}}$}\xspace}
\newcommand{\roc}{\mbox{$\mathop{\mathtt{ROC}}$}\xspace}
\newcommand{\rocfive}{\mbox{$\mathop{\mathtt{ROC}_{\scriptsize{5}}}$}\xspace}
\newcommand{\rocten}{\mbox{$\mathop{\mathtt{ROC}_{\scriptsize{10}}}$}\xspace}
\newcommand{\blosum}{\mbox{$\mathop{\mathtt{BLOSUM}}\limits$}\xspace}
\newcommand{\onehot}{\mbox{$\mathop{\mathtt{Onehot}}\limits$}\xspace}
\newcommand{\deep}{\mbox{$\mathop{\mathtt{Deep}}\limits$}\xspace}
\newcommand{\wordtovec}{\mbox{$\mathop{\mathtt{Word2Vec}}\limits$}\xspace}

\newcommand{\vect}[1]{\boldsymbol{#1}}
\newcommand{\matr}[1]{\uppercase{#1}}
\newcommand{\R}{\mathbb{R}}
\newcommand{\metric}{\mbox{$\mathop{\text{mtrc}}\limits$}\xspace}

\newcommand{\allele}{\mbox{$\mathop{Q}$}\xspace}
\newcommand{\pset}{\mbox{$\mathop{\mathcal{P}}$}\xspace}
\newcommand{\peptide}{\mbox{$\mathop{p}$}\xspace}
\newcommand{\peptidei}{\mbox{$\mathop{p}_i$}\xspace}
\newcommand{\amino}{\mbox{$\mathop{a}\limits$}\xspace}
\newcommand{\nbf}{\mbox{$\mathop{b}\limits$}\xspace}
\newcommand{\lev}{\mbox{$\mathop{l}\limits$}\xspace}
\newcommand{\kmer}{\mbox{$\mathop{k}\limits$}\xspace}
\newcommand{\entry}{\mbox{$\mathop{m}\limits$}\xspace}

\newcommand{\Sf}{\mbox{$\mathop{S}\limits$}}
\newcommand{\ranki}{\mbox{$\mathop{s}_i$}}
\newcommand{\thd}{\mbox{$\mathop{h}\limits$}\xspace}
\newcommand{\nthd}{\mbox{$\mathop{t}\limits$}\xspace}
\newcommand{\hybrid}{\mbox{$\mathop{H}\limits$}\xspace}

\newcommand{\aenc}{\mbox{$\mathop{\vect{e}}\limits$}\xspace}
\newcommand{\vhd}{\mbox{$\mathop{\vect{r}}\limits$}\xspace}
\newcommand{\mhd}{\mbox{$\mathop{\matr{R}}\limits$}\xspace}
\newcommand{\vathd}{\mbox{$\mathop{\vect{c}}\limits$}\xspace}
\newcommand{\dd}{\mbox{$\mathop{d_f}\limits$}\xspace}
\newcommand{\dpos}{\mbox{$\mathop{d_o}\limits$}\xspace}
\newcommand{\dhd}{\mbox{$\mathop{d_r}\limits$}\xspace}
\newcommand{\dathd}{\mbox{$\mathop{d_a}\limits$}\xspace}
\newcommand{\daenc}{\mbox{$\mathop{d_e}\limits$}\xspace}
\newcommand{\pos}{\mbox{$\mathop{\vect{o}}\limits$}\xspace}
\newcommand{\vfea}{\mbox{$\mathop{\vect{f}}\limits$}\xspace}

\newcommand{\icfif}{\mbox{$\text{IC}_{50}$}\xspace}

\newcommand{\xia}[1]{\textcolor{red}{#1}}
\newcommand{\ziqi}[1]{\textcolor{blue}{#1}}

\def\keyFont{\fontsize{8}{11}\helveticabold }
\def\firstAuthorLast{Chen {et~al.}} %use et al only if is more than 1 author
\def\Authors{Ziqi Chen\,$^{1}$, Martin Renqiang Min\,$^{2}$ and Xia Ning\,$^{1,3,4}$}
\def\Address{$^{1}$Computer Science and Engineering Department, The Ohio State University, Columbus, OH, US\\
$^{2}$Machine Learning Department, NEC Labs America, Princeton, New Jersey, NJ, US\\
$^{3}$Biomedical Informatics Department, The Ohio State University, Columbus, OH, US\\
$^{4}$Translational Data Analytics Institute, The Ohio State University, Columbus, OH, US
}
\def\corrAuthor{Xia Ning}
\def\corrEmail{ning.104@osu.edu}

\begin{document}
\onecolumn
\firstpage{1}

%===============================================================
\title[RCNN4PMHC]{Ranking-based Convolutional Neural Network Models for Peptide-MHC Binding Prediction}
%===============================================================

\author[\firstAuthorLast ]{\Authors} %This field will be automatically populated
\address{} %This field will be automatically populated
\correspondance{} %This field will be automatically populated

\extraAuth{}% If there are more than 1 corresponding author, comment this line and uncomment the next one.
%\extraAuth{corresponding Author2 \\ Laboratory X2, Institute X2, Department X2, Organization X2, Street X2, City X2 , State XX2 (only USA, Canada and Australia), Zip Code2, X2 Country X2, email2@uni2.edu}

\maketitle

%%%%%%%%%%%%%%%%%%%%%%%%%%%%%%%%%%%%%%%%%%%%%%%%%%%%%%%%%
\begin{abstract}
%%%%%%%%%%%%%%%%%%%%%%%%%%%%%%%%%%%%%%%%%%%%%%%%%%%%%%%%%

%\section{}
T-cell receptors can recognize foreign peptides bound to major histocompatibility complex (MHC) class-I proteins, 
and thus trigger the adaptive immune response.
Therefore, identifying peptides that can bind to MHC class-I molecules plays a vital role in the design of peptide vaccines.
Many computational methods, for example, the state-of-the-art allele-specific method \mhcflurry, have been developed to predict 
the binding affinities between peptides and MHC molecules.
In this manuscript, we develop two allele-specific Convolutional Neural Network (CNN)-based methods 
named \convmodel and \spannyconvmodel to tackle the binding prediction problem.
Specifically, we formulate the problem as to optimize the rankings of peptide-MHC bindings via ranking-based learning 
objectives. 
%Specifically, we develop three ranking-based learning objectives to learn the ordering of peptide-MHC pairs.
%
%These new learning objectives also enable our models to be 
Such optimization is 
more robust and tolerant to the measurement inaccuracy of binding affinities, and therefore enables more 
accurate prioritization of binding peptides.
In addition, we develop a new position encoding method in \convmodel and \spannyconvmodel to 
better identify the most important amino acids for the binding events.
%
%We also apply various ranking-based metrics to evaluate the performance of our models.
%
We conduct a comprehensive set of experiments using the latest Immune Epitope Database (IEDB) datasets.
Our experimental results demonstrate that our models significantly outperform the state-of-the-art
methods including \mhcflurry with an average percentage improvement of 6.70\% on AUC and 
17.10\% on ROC5 across 128 alleles. 
%and that the ranking-based optimization is able to better prioritize binding peptides.

\tiny
\keyFont{ \section{Keywords:} deep learning, prioritization, peptide vaccine design, convolutional neural networks, attention} %All article types: you may provide up to 8 keywords; at least 5 are mandatory.
\end{abstract}

%%%%%%%%%%%%%%%%%%%%%%%%%%%%%%%%%%%%%%%%%%%%%%%%%%%%%%%%%
\section{Introduction}
\label{sec:intro}
%%%%%%%%%%%%%%%%%%%%%%%%%%%%%%%%%%%%%%%%%%%%%%%%%%%%%%%%%

Immunotherapy, an important treatment of cancers, treats the disease by boosting patients' immune systems to 
kill cancer cells~\cite{Waldman2020,Esfahani2020,Couzin-Frankel1432,Mellman2011}.
To trigger patients' adaptive immune responses, Cytotoxic T cells, also known as CD8+ T-cells, have to recognize 
peptides presented on the cancer cell surface~\cite{Blum2013, valitutti1995serial}.
These peptides are fragments derived from self-proteins or pathogens by proteasomal proteolysis within the cell.
%If a peptide presented on the cell surface is from pathogen and recognized by TCRs, Cytotoxic T cells will trigger 
%the adaptive immune response to kill the cell \cite{Blum2013}.
%
To have the peptides presented on the cell surface to be recognized by CB8 receptors, 
%
%Note that peptides that are recognized by CD8 receptors, have to be presented on the cell surface.  
%
they need to be brought from inside the cells to the cell surface, typically through binding with and transported by 
major histocompatibility complex (MHC) class-I molecules. 
%
%Major histocompatibility complex (MHC) class-I molecules can bind to peptides and bring them to the cell surface.
%
%
%To help \hl{CD8} receptor recognize \hl{pathogen}, % and trigger immune responses, 
%\hl{peptide vaccines have been being developed by mimicking natural proteins from pathogens}~\cite{Purcell2007}.
%
To mimic natural occurring proteins from pathogens, synthetic peptide vaccines are developed 
for therapeutic purposes~\cite{Purcell2007}. 
Therefore, to design successful peptide vaccines, it is critical to identify and study peptides that can bind with MHC 
class-I molecules.
%
%Major histocompatibility complex (MHC) class-I genes (or alleles) encode MHC class-I molecules.
%
%Peptides can bind to MHC class-I molecules, and be brought to the outer cell surface by the MHC class-I molecule for the recognition of TCRs.
%

Many computational methods have been developed to predict the binding affinities between peptides and MHC class-I 
molecules~\cite{Donnell2018,Han2017}.
These existing computational methods can be categorized into two types: allele-specific methods and pan methods. 
Allele-specific methods train one model for one allele such that the model can capture binding patterns specific to the allele, and thus it is better 
customized to that allele~\cite{Lundegaard2008,Donnell2018}.
Pan methods train one model for all the alleles at a same time, and thus the information across different alleles can be shared and
integrated into a general model~\cite{Jurtz2017,Hu2018}.
These existing methods can achieve significant performance on the prediction of binding affinities.
However, most existing methods formulate the prediction problem as to predict the exact binding affinity values (e.g., IC$_{50}$ values) via regression. 
Such formulations may suffer from two potential issues. First of all, they tend to be sensitive to the measurement errors when the measured IC$_{50}$
values are not accurate. In addition, many of these methods use ranking-based measurement such as Kendall's Tau correlations 
to measure the performance of regression-based methods~\cite{Bhattacharya2017,Odonnell2020}. 
This could lead to sub-optimal solution as small regression errors do not necessarily correlate 
to large Kendall's Tau. 
%
%Since prioritization (ranking) of peptides in terms of their binding activities 
%% 
%%match with their learning objective for regression problem (\cite{Bhattacharya2017} and \cite{Odonnell2020}).
%%
%%In order to incorporate self-attention layer to our method, we propose a new position encoding method to distinguish peptides 
%%of different positions. 
Therefore, these methods are limited in their capability of prioritizing the most possible peptide-MHC pairs of high binding affinities.

In this study, we formulate the problem as to prioritize the most possible peptide-MHC binding pairs via ranking based learning. 
%
%To prioritize more promising peptide-MHC pairs, we formulate the problem as to optimize the ranking of peptide-MHC pairs.
%
We propose three ranking-based learning objectives %, denoted as \hingeone, \hingetwo and \hingethree, 
such that through optimizing these objectives, we impose peptide-MHC pairs of high binding affinities ranked higher than 
those of low binding affinities.
Coupled with these objectives, we develop two allele-specific Convolutional Neural Network (CNN)-based methods 
with attention mechanism, denoted as \convmodel and \spannyconvmodel.
\convmodel extracts local features of peptide sequences using 1D convolutional layers, and learns the importance of different positions in peptides
using self-attention mechanism.
In addition to the local features used in \convmodel, 
%\spannyconvmodel also incorporates global features 
%of the entire peptide sequences. Therefore, 
%
\spannyconvmodel represents the peptide sequences at different granularity levels by leveraging both global and local features 
of peptide sequences.
We also develop a new position encoding method together with self-attention mechanism so as to differentiate amino acids at different positions.
We compare the various combinations of model architectures and objective functions of our methods with the 
state-of-the-art baseline \mhcflurry \cite{Donnell2018} on IEDB datasets~\cite{Vita2018}.
Our experimental results demonstrate that our models significantly outperform the state-of-the-art
methods with an average percentage improvement of 6.70\% on AUC and 
17.10\% on ROC5 across 128 alleles. 
%
%Specifically, our \spannyconvmodel model with \hingeone loss achieves 6.70\% improvement on \auc over the baseline.
%Our \convmodel model with \hingeone loss achieves 5.13\% improvement on \auc over the baseline.
%%

We summarize our contributions below:
\begin{itemize}
	\item
	We formulate the problem as to optimize the rankings of peptide-MHC pairs instead of predicting the exact binding affinity values. 
	Our experimental results demonstrate that our ranking-based learning is able to significantly improve the performance of 
	identifying the most possible peptide-MHC binding pairs.
	\item
	We develop two allele-specific methods \convmodel and \spannyconvmodel with position encoding and self attention, which 
	enable a better learning of the importance of amino acids at different positions in determining peptide-MHC binding. 
	\item 
	We incorporate both global and local features in \spannyconvmodel to better capture and learn from different granularities of 
	peptide sequence information.  
	%
%	\item
%	The self-attention layer in \convmodel can learn the importance of amino acids of different
%	positions to the binding events of peptide-MHC pairs. 
	\item
	Our methods outperform the state-of-the-art baseline \mhcflurry on IEDB datasets~\cite{Donnell2018} in prioritizing the most 
	possible peptide-MHC binding pairs.
\end{itemize}
%

%%%%%%%%%%%%%%%%%%%%%%%%%%%%%%%%%%%%%%%%%%%%%%%%%%%%%%%%%%%%%%%%%%%%%%%%%%%%%%%%%%%%
\section{Literature Review} 
\label{sec:review}
%%%%%%%%%%%%%%%%%%%%%%%%%%%%%%%%%%%%%%%%%%%%%%%%%%%%%%%%%%%%%%%%%%%%%%%%%%%%%%%%%%%%

The existing computational methods for peptide-MHC binding prediction can be generally classified into two 
categories: linear regression-based methods and deep learning (DL)-based methods. Below, we present a literature 
review for each of the categories, including the key ideas and the representative work. 

%***********************************************************************************
\subsection{Peptide Binding Prediction via Linear Regression}
\label{sec:review:LR}
%************************************************************************************

%
Many early developed methods on peptide-MHC binding prediction are based on linear regression.
For example, Peters \etal~\cite{Peters2005} proposed a method named Stabilized Matrix Method 
({\SMM}), which applied linear regression to predict the binding affinities from one-hot encoded 
vector representation of peptide sequences.
Kim \etal~\cite{Kim2009} derived a novel amino acid similarity matrix named Peptide:MHC Binding Energy 
Covariance (PMBEC) matrix and incorporated it into the \SMM approach to improve the performance of 
\SMM. In PMBEC, each amino acid is represented by its 
covariance of relative binding energy contributions with all other amino acids.
Some recent work~\cite{Bonsack2019, Zhao2018} demonstrates these linear regression-based methods are
inferior to DL-based methods, 
%such as NetMHCpan~\cite{Jurtz2017} outperform these linear regression-based methods in general.
%
and therefore, in our work, we focus on DL-based methods.

%***********************************************************************************
\subsection{Peptide Binding Prediction via Deep Learning}
\label{sec:review:DL}
%************************************************************************************

The DL-based models can be categorized into allele-specific methods and pan methods.
Allele-specific methods train a model for each allele and learn the binding patterns of each allele separately.
Instead, pan methods train a model for all alleles to learn all the binding patterns together within one model. 
Both the methods use similar encoding methods such  \onehot encoding, \blosum
encoding and \wordtovec~\cite{Goldberg2014}.
%
%Most NN-based models on peptide binding prediction primarily differ in feature learning, and 
%after feature learning, most models would apply fully connected layer to predict the binding probability.

%
%------------------------------------------------------------------------------------
\subsubsection{Allele-Specific Deep Learning Methods}
\label{sec:review:DL:allele}
%------------------------------------------------------------------------------------

Among these allele-specific methods, Lundegaard \etal~\cite{Lundegaard2008} proposed \NetMHCtrd that takes
the embeddings of peptide sequences as input, and they 
applied neural networks with one hidden layer to predict peptide-MHC binding for peptides of fixed length.
In \NetMHCtrd, the hidden layer is a fully-connected (FC) layer, and learns the 
global features of peptide sequences such as the position and types of specific amino acids.
Andreatta \etal~\cite{Andreatta2015} extended \NetMHCtrd to \NetMHCfth by padding  
 so that the model can handle peptides of variable length.
Kuksa \etal~\cite{Kuksa2015} developed two 
nonlinear high-order methods including high-order neural networks (\HONN) pre-trained with high-order semi-restricted 
Boltzmann machine (RBM), and high-order kernel support vector machines (\hkSVM). 
Both the high-order RBMs and the high-order kernel are designed to capture the direct strong 
high-order interactions between features.
Bhattacharya \etal~\cite{Bhattacharya2017} developed a deep recurrent neural network 
based on gated recurrent units (GRUs) to
capture the sequential features from peptides of various length. 
Vang \etal~\cite{Vang2017} applied two layers of 1D convolution on the embeddings of peptide sequences
so as to learn local binding patterns existing in each $k$-mer amino acids.
O'Donnell \etal~\cite{Donnell2018} designed a deep model named \mhcflurry with locally-connected layers.
This locally-connected layer is used to learn the position-specific local features from 
the peptide sequences. %, which to some extent combines the fully connected layers and convolutional layers.
MHCflurry has been demonstrated to achieve better or similar performance compared with  most of 
The other prediction methods~\cite{Boehm2019, Zhao2018}.

%------------------------------------------------------------------------------------
\subsubsection{Pan Deep Learning Methods}
\label{sec:review:DL:pan}
%------------------------------------------------------------------------------------

%As for pan methods,
%
Nielsen \etal \cite{Nielsen2016} developed a DL-based pan method named NetMHCpan3.0.
This method takes the embedding of pseudo MHC sequences and peptide sequences as input, and then applies an ensemble of neural networks to 
predict the binding affinities of peptide-MHC pairs.
Jurtz \etal~\cite{Jurtz2017} extended NetMHCpan3.0 to NetMHCpan4.0 by training the model on both 
binding affinity data and eluted ligand data. 
Their model shares a hidden layer among two kinds of data and applies two different output layers to predict binding 
affinities and eluted ligands, respectively, for peptide-MHC pairs.
Phloyphisut \etal~\cite{Phloyphisut2019} developed a deep learning model, which uses 
GRUs to learn the embeddings of peptides, and a FC layer to learn the embeddings 
of alleles. The two types of embeddings are then concatenated to predict peptide-MHC binding probabilities.
Han \etal~\cite{Han2017} encoded peptide-MHC pairs into image-like array (ILA) data and applied 
deep 2D convolutional neural networks to extract the possible peptide-MHC interactions from the ILA data.  
Hu \etal~\cite{Hu2018} combined a deep convolutional neural network with an attention module. 
They applied multiple convolutional layers to extract features of different levels. 
The extracted features are integrated with the features learned from attention mechanism 
and fed into the output layer to predict binding affinities of peptide-MHC pairs.
O'Donnel \etal~\cite{Odonnell2020} developed a pan-allele binding affinity predictor \mhcflurrynd BP and an allele-independent antigen 
presentation predictor \mhcflurrynd AP to calculate the presentation scores of peptide-MHC pairs.
Their binding affinity predictor includes upstream and downstream residues of peptides from their source proteins to improve 
the performance of models.
Note that \mhcflurrynd is a pan method and requires source proteins of peptides.
Therefore, we do not compare our methods with \mhcflurrynd.

%%%%%%%%%%%%%%%%%%%%%%%%%%%%%%%%%%%%%%%%%%%%%%%%%%%%%%%%%%%%%%%%%%%%%%%%%%%%%%%%%%%%
\section{Materials}
\label{sec:datasets}
%%%%%%%%%%%%%%%%%%%%%%%%%%%%%%%%%%%%%%%%%%%%%%%%%%%%%%%%%%%%%%%%%%%%%%%%%%%%%%%%%%%%

%************************************************************************************
\subsection{Peptide-MHC Binding Data}
\label{sec:datasets:iedb}
%************************************************************************************
%
The dataset is collected from the Immune Epitope Database (IEDB) \cite{Vita2018}. 
Each peptide-MHC entry \entry 
in the dataset measures the binding affinity between a peptide and an allele.
These binding affinity entries could be of either quantitative values (e.g., \icfif) or qualitative levels indicating levels of binding strength.
The mapping between quantitative values and qualitative levels is shown in Table~\ref{tbl:level}.
Note that higher $\icfif$ values indicate lower binding affinities. 

\begin{table}[!h]
	\begin{center}
		\caption{Binding Affinity Measurement Mapping}
		\label{tbl:level}
		\centering
		\begin{threeparttable}
			%    \begin{small}
			\begin{tabular}{
					@{\hspace{5pt}}l@{\hspace{5pt}}r@{\hspace{5pt}}r@{\hspace{5pt}}
					@{\hspace{5pt}}l@{\hspace{5pt}}r@{\hspace{5pt}}r@{\hspace{5pt}}       
				}
				\toprule
				qualitative & quantitative & level\\
				\midrule
				negative   & >5,000nM & 1\\
				positive-low   & 1,000-5,000nM & 2\\
				positive-intermediate   & 500-1,000nM & 3\\
				positive    & 100-500nM & 4\\
				positive-high & 0-100nM & 5\\
				
				\bottomrule
			\end{tabular}
			%    \end{small}
			%\begin{tablenotes}
			%\item \par
			%\end{tablenotes}
		\end{threeparttable}
	\end{center}
\end{table}
%\label{tbl:level}

We combined the widely used IEDB benchmark dataset curated by Kim et al. \cite{Kim2014} and
the latest data added to IEDB (downloaded from the IEDB website on Jun. 24, 2019).
The benchmark dataset contains two datasets BD2009 and BD2013 
%that are two MHC-I binding data files 
compiled in 2009 and 2013, respectively.
BD2009 consists of 137,654 entries, and BD2013 consists of 179,692 entries.
The latest dataset consists of 189,063 peptide-MHC entries.
Specifically, we excluded those entries with non-specific, mutant or unparseable allele names such as HLA-A2.
We then combined the datasets by processing the duplicated entries and entries with conflicting affinities as follows. %
We first mapped the quantitative values of all these duplicated or conflicting entries into qualitative levels based on Table~\ref{tbl:level}, 
and used majority voting to identify the major binding level of the peptide-MHC pairs.
%
%If we can specify the unique major binding level for a peptide-MHC pair, we set the binding affinity value of that pair as the average of quantitative binding affinity values 
%belonging to that major binding level; otherwise, we simply removed all the conflicting entries for that pair.
If such binding levels cannot be identified, we simply removed all the conflicting entires;
otherwise, we assigned the average quantitative values in the identified major binding level to the peptide-MHC pairs.
The combined dataset consists of 202,510 entries across 128 alleles and 53,253 peptides as in Table~\ref{tbl:stat}. 
%
%Considering that the measured quantitative binding affinity values ranging from 0 to $10^7$ are difficult for model to predict, 
We further normalized the binding affinity values ranging from 0 to $10^7$ to $[$0, 1$]$ via formula 
\begin{equation}
\label{eqn:clamp}
\nbf=\text{clamp}(1-\log_{50,000}(x), 0, 1), 
\end{equation}
where $x$ is the measured binding affinity value, and $\text{clamp}(1-\log_{50,000}(x), 0, 1)$ represents that \mbox{$1-\log_{50,000}(x)$} 
is clamped into range $[0, 1]$.
By using the above clamp function, smaller/larger binding affinity values corresponding to higher/lower binding affinities will be converted to higher/lower
 normalized values. 

\begin{table}[!h]
	\caption{Data Statistics}
	\label{tbl:stat}
	\centering
	\begin{threeparttable}
		%    \begin{small}
		\begin{tabular}{
				@{\hspace{10pt}}l@{\hspace{10pt}}
				@{\hspace{10pt}}r@{\hspace{10pt}}
			}
			\toprule
			variables & count\\
			\midrule
			entries & 202,510\\
			alleles & 128 \\
			peptides & 53,253\\
			\bottomrule
		\end{tabular}
		%    \end{small}
		%\begin{tablenotes}
		%\item \xia{what do yo mean "positive" and "negative" peptides?}
		%\end{tablenotes}
	\end{threeparttable}
\end{table}
%

%%%%%%%%%%%%%%%%%%%%%%%%%%%%%%%%%%%%%%%%%%%%%%%%%%%%%%%%%%%%%%%%%%%%%%%%%%%%%%%%%%%%
\section{Definitions and Notations}
\label{sec:notations}
%%%%%%%%%%%%%%%%%%%%%%%%%%%%%%%%%%%%%%%%%%%%%%%%%%%%%%%%%%%%%%%%%%%%%%%%%%%%%%%%%%%%
%
All the key definitions and notations are listed in Table~\ref{tbl:notation}. 

\begin{table}[!h]
	\caption{Notations}
	\label{tbl:notation}
	\centering
	\begin{threeparttable}
		\begin{tabular}{
				@{\hspace{2pt}}l@{\hspace{5pt}}
				@{\hspace{2pt}}l@{\hspace{2pt}}
				@{\hspace{2pt}}l@{\hspace{2pt}}          
			}
			\toprule
			&	notation & meaning \\
			\midrule
			\multirow{4}{*}{peptides \& alleles}
			& \peptide & a peptide\\
			& \amino & an amino acid of a peptide sequence\\
			& \pset   & a set of peptides\\
			& \allele & an allele\\
			\midrule
			\multirow{2}{*}{binding} 
			& $x$/\nbf   & original/normalized binding affinity for a peptide-MHC pair\\
			& \lev  & binding level for a peptide-MHC pair\\
			\midrule
			\multirow{6}{*}{embeddings}
			& \aenc  & encoding vector of amino acid type\\
			& \vhd   & embedding vector of each amino acid \\
			& \pos   & position embedding of each $k$-mer amino acids\\
			& \mhd   & feature matrix for a peptide sequence \\
			& $\matr{F_g}$   & feature matrix for a padded peptide sequence \\
			&                & (i.e., input of global kernel in \spannyconvmodel) \\
			\midrule
			\multirow{8}{*}{{parameters}}
			& $\Sf(\cdot)$    & a scoring function \\
			& \daenc  & dimension of amino acid embedding \\
			& \dd  & number of filters in convolution layer\\
			& \dpos  & dimension of position embedding \pos\\
			& $d_g$   & number of global kernels in \spannyconvmodel \\
			& \dhd  & dimension of hidden units \vathd in fully connected layer\\
			& \kmer   & kernel size in convolutional neural layer   \\
			& $w$     & attention value learned in attention layer\\
			%\E   & embedding matrix for peptide \\
			\bottomrule
		\end{tabular}
	\end{threeparttable}
\end{table}
%
%  \label{tbl:notation}

%%%%%%%%%%%%%%%%%%%%%%%%%%%%%%%%%%%%%%%%%%%%%%%%%%%%%%%%%%%%%%%%%%%%%%%%%%%%%%%%%%%%
\section{Methods}
\label{sec:methods}
%%%%%%%%%%%%%%%%%%%%%%%%%%%%%%%%%%%%%%%%%%%%%%%%%%%%%%%%%%%%%%%%%%%%%%%%%%%%%%%%%%%%

%We study the models based on their model architectures and loss functions, and thus all the model-loss function combinations. 
%
We developed two new models: \convmodel and \spannyconvmodel (will be discussed in Section~\ref{sec:methods:model:convmodel} and 
Section~\ref{sec:methods:model:spannyconvmodel}), and compare them with \mhcflurry \cite{Donnell2018}, where \mhcflurry is the state-of-the-art and used as the baseline. 
In terms of the embeddings of amino acids, we compare the performance of \spannyconvmodel with three embedding methods for amino acids and their combinations.
In terms of the loss functions, we developed three pair-wise hinge loss functions,
and compare them with the conventional mean-square loss function used in \mhcflurry.

%====================================================================================
\subsection{Convolutional Neural Networks with Attention Layers (\convmodel)}
\label{sec:methods:model:convmodel}
%====================================================================================

\begin{figure}[!t] 
	\centering
	\includegraphics[width=\linewidth]{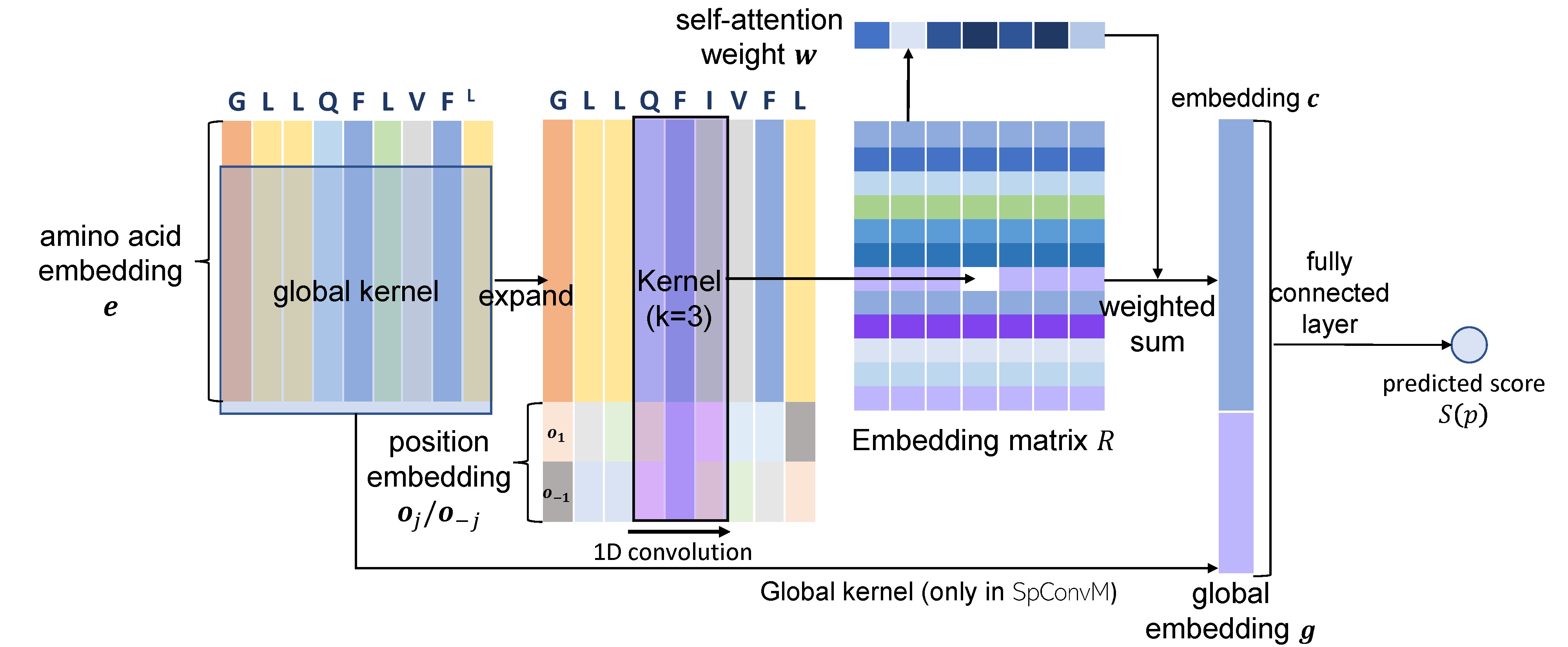}
	\caption{Architectures of \convmodel and \spannyconvmodel}
	\label{fig:conv}
	\vspace{5pt}
\end{figure}

In this section, we introduce our new model \convmodel, a convolutional neural network with attention layers. 
Figure~\ref{fig:conv} presents the architecture of \convmodel. 
%

%------------------------------------------------------------------------------------------------------
\subsubsection{Peptide Representation in \convmodel}
%------------------------------------------------------------------------------------------------------
%
%We developed a deep learning-based model, denoted as \convmodel. 
%
In \convmodel, we first represent each amino acid, 
denoted as $\amino_j$, in a peptide sequence, denoted as $\peptide=[\amino_1,...,\amino_j,...,\amino_n]$ (started from C-end), using two 
types of information. 
The first information encodes the type of amino acids using BLOSUM62 matrix or \onehot encoding. 
The details about the encoding of amino acids are 
described in Section~\ref{sec:experiments:baseline:encoding}. %\xia{please briefly describe the information here}
The second information encodes the position of each amino acid in peptide sequences, as it has been demonstrated 
\cite{Donnell2018} that different positions of peptides contribute differently to their binding to alleles. 
%
%We randomly initialize a position embedding matrix $\pset \in R^{l_{max}\times d_p}$, in which 
%$l_{max}$ is the maximum length of peptides and $d_p$ is the dimension of position embeddings.
%
In particular, each position of the peptide sequences, regardless of 
which amino acid is at the position, will have two position vectors: for the $j$-th position from the C-end, 
we use {$\pos_j$ and $\pos_{-j}\in \R^{\dpos\times 1}$} to represent the position information with respect to the C-end and 
the N-end, respectively. 
The two position vectors will together accommodate the variation of peptide lengths. 
Thus, each amino acid $\amino_j$ is represented as a feature vector 
$\vfea_j=[\aenc_j;\pos_j;\pos_{-j}]\in  \R^{(\daenc+2\dpos)\times 1}$, 
where {$\aenc_j\in \R^{\daenc\times 1}$} 
%\xia{you need to double check all the vector representations in the entire paper.. 
%by default it should be row vectors or column vectors? Here the representation does not match the dimensionality... }
%\ziqi{I already carefully double-checked the dimension here...maybe we don't need to show the dimensionality here, 
%most papers don't show the dimensionality of their parameters...}
is $\amino_j$'s embedding with an encoding method in Section~\ref{sec:experiments:baseline:encoding}, 
%\xia{unify Sec to Section through out the entire paper}
and $\dpos$ is the dimension of position embedding vectors; 
%$\vect{p}^j/\vect{p}^{-j} \in \R^{1\times d_p}$ \xia{dimension is not correct!} 
%Is a vector for the position of the amino acid to the C-end and to the 
%N-end, respectively. 
%
a peptide of $n$ amino acids is represented as a feature matrix 
{$\matr{F} = [\vfea_1,\vfea_2,...,\vfea_n]\in \R^{(\daenc+2\dpos)\times n}$.}
%\xia{again, the representation does not match the dimensionality... }
%\xia{here $m$ is abused since it was used for Peptide-MHC pairs before... }.
%
%Note that the position vectors are determined by the positions of the amino acids, not by the amino acids themselves, and 
%therefore different amino acids may have same position vectors if they are at the same position in two different peptides.
%
%\st{With two position values, we obtain two position vectors $p_0\in \R^{d_p},p_1\in \R^{d_p}$ by 
%looking up the shared matrix $P$.
%%%
%The final position embedding $p$ is obtained by concatenating two position vectors $p=(p_0, p_1)\in \R^{2d_p}$.
%%
%We concatenate the position embedding $p$ with the embedding $a$ of the amino acid and get the 
%feature representation $f=(p,a)\in \R^{20+2d_p}$ for each amino acid in the sequence. 
%%
%Consequently, a peptide of length $l$ will be encoded into a feature matrix 
%$F=(f_1,f_2,...,f_l)\in \R^{l \times (20+2d_p)}$.}
%
The position vectors will be learned in order to optimize the peptide representations.

%
%------------------------------------------------------------------------------------------------------
\subsubsection{Model Architecture of \convmodel}
%------------------------------------------------------------------------------------------------------
%
The \convmodel model consists of a 1D convolutional layer, a self-attention layer and 
a fully-connected layer as demonstrated in Figure~{\ref{fig:conv}}.
The 1D convolutional layer takes the peptide feature matrix \mbox{$\matr{F}\in \R^{(\daenc+2\dpos)\times n}$} as input, and extracts local 
feature patterns from the peptide sequences via 1D convolution using $\dhd$ kernels of size $(\daenc+2\dpos)\times k$.
The output of the 1D convolutional layer is an embedding matrix \mbox{$\mhd = [\vhd_1,...,\vhd_{(n-k+1)}]\in \R^{\dhd\times (n-k+1)}$}, 
in which each column $\vhd_i$ 
%\xia{dimensionality is WRONG! Please carefully check all the matrix/vector representations through out the entire appear...}
represents the embedding of the $i$-th $k$-mer 
out of the \dhd kernels.
Batch normalization is applied to fix the mean and variance of the embedding matrix $\mhd$ in each batch.
After batch normalization, rectified linear unit (ReLU) activation is applied as the non-linear activation function on the 
embedding matrix.
%
%On the $k$-mer embedding matrix $E$,
Then, we apply the self-attention mechanism \cite{Chorowski2015}
%, which has been widely used to improve the interpretability of deep models,
%
to convert the embedding matrix into an embedding vector $\vathd$ for the input peptide as follows.
First, the weight $w_i$ for $i$-th $k$-mer is calculated as follows, 
\begin{equation}
\label{eqn:att}
w_i = \frac{\exp(a_i)}{\sum_j \exp(a_j)}, ~~
{a_i = \vect{v}\tanh(\matr{W}\vhd_i+\vect{b})}, 
\end{equation}
where {$\matr{W}\in \R^{\dathd\times \dhd}$}, $\vect{b}\in \R^{\dathd\times 1}$ and $\vect{v}\in \R^{1\times \dathd}$ 
are the parameters of self-attention layer, 
and $\dathd$ is the number of hidden units in the self-attention layer.
%
%Equation~\ref{eqn:att} maps the embedding matrix $E$ into a weight vector $\vect{a}$,
%%
%then the softmax function is applied to make the values of attention weights sum up to 1.
%
With the weight $w_i$ on each $k$-mer, the embedding of the whole sequence is calculated as the weighted 
sum of all $k$-mer embeddings, that is,
\begin{equation}
\label{eqn:sum}
\vathd=\sum_{i}^{n-k+1}{w_i \vhd_i}. 
\end{equation}
The embedding vector $\vathd\in \R^{\dhd\times 1}$ of the input peptide is then fed into the fully-connected layer to predict 
peptide binding at the output layer. 
We will discuss the loss function used at the output layer later in Section~\ref{sec:methods:loss}.
%\clearpage

%====================================================================================
\subsection{Convolutional Neural Networks with Global Kernels and Attention Layers (\spannyconvmodel)}
\label{sec:methods:model:spannyconvmodel}
%====================================================================================

We further develop \convmodel into a new model with global kernels, denoted as \spannyconvmodel as in 
Figure~\ref{fig:conv}.
The use of global kernels is inspired by Bhattacharya \etal~\cite{Bhattacharya2017}, which 
demonstrates that global kernels within CNN models can significantly improve the performance of peptide 
binding prediction.
As \convmodel primarily extracts and utilizes local features, the additional global kernels extract 
global features from the entire peptide sequences that could be useful for prediction but cannot be captured 
by local convolution.
In order to use global kernels, we pad the peptide sequences of various lengths to length 15, 
with padding 0 vectors  in the middle of the peptide representations in a same way as in \mhcflurry.
More details about padding are available later in Section~\ref{sec:experiments:baseline:mhcflurry}. 
The padded peptide sequences will be encoded into a feature matrix $\matr{F_g}$
in the same way as in \convmodel, except that the position embeddings are not included
because the global kernels will overwrite the local information after the convolution.\\

Given the input $\matr{F_g}$, the convolution using $d_g$ global kernels will generate a vector $\vect{g}\in \R^{d_g\times 1}$. 
We concatenate $\vect{g}$ and $\vathd$ as in \convmodel %(Equation~\ref{eqn:sum})  
(i.e., the embedding vector calculated from local kernels)
to construct a local-global embedding vector $\vathd'=[\vathd;\vect{g}]$ for the input peptide 
sequence and feed $\vathd'$ into the fully-connected layer to predict peptide prediction 
as in Figure~\ref{fig:conv}. 
%

%************************************************************************************
\subsection{Loss Functions}
\label{sec:methods:loss}
%************************************************************************************

%In order to convert the original binding prediction problem into a new ranking problem, 
We propose three pair-wise hinge loss functions, denoted as \hingeone, \hingetwo and \hingethree, respectively. 
We will compare these loss functions with the widely used mean-square loss function~\cite{Donnell2018}, 
denoted as \meansquare, in learning peptide bindings.

%====================================================================================
\subsubsection{Hinge Loss Functions for Peptide Binding Ranking}
\label{sec:methods:loss:hingeloss}
%====================================================================================

%
We first evaluate the hinge loss as the loss function in conjunction with various model architectures.
The use of hinge loss is inspired by the following observation.
We noticed that in literature, peptide-MHC binding prediction is often formulated into either a regression 
problem, in which the binding affinities between peptides and alleles are predicted, or a classification problem, 
in which whether the peptide will bind to the allele is the target to predict.
However, in practice, it is also important to first prioritize the most promising peptides with acceptable binding 
affinities for further assessment, whereas regression and classification are not optimal for prioritization. 
Besides, recent work has already employed several evaluation metrics on top ranked peptides, for example, 
\cite{Zeng2019} evaluated the performance through the true positive rate at 5\% false positive rate, 
%\xia{what does this mean?}
%\ziqi{it is similar to ROC5/ROC10 and refers to the rate of true-positive or the percentage of true-positive at certain false positive rate}
which suggests the importance of top-ranked peptides in addition to accurate affinity prediction.
All of these inspire us to consider ranking based formulation for peptide prioritization.

Given two normalized binding affinity values $\nbf_i$ and $\nbf_j$ of any two peptides $\peptide_i$ and $\peptide_j$
with respect to an allele, 
the allele-specific pair-wise ranking problem can be considered as to learn a scoring function
{$\Sf(\cdot)$}, such that
\begin{equation}
\label{eqn:score}
\Sf(\peptide_i)>\Sf(\peptide_j), \text{ if }\ \nbf_i > \nbf_j. 
\end{equation}
Please note that $\Sf(\peptide_i)$ is a score for peptide $\peptide_i$, which is not necessarily close to the binding affinity 
$\nbf_i$, as long as it reconstructs the ranking structures among all peptides.
This allows the ranking based formulation more flexibility to identify the most promising peptides without accurately estimating 
their binding affinities.
To learn such scoring functions, hinge loss is widely used, and thus we develop three hinge loss functions to emphasize 
different aspects during peptide ranking.
%

%------------------------------------------------------------------------------------------------------
\paragraph{Value-based Hinge Loss Function}
%------------------------------------------------------------------------------------------------------
%
%
The first hinge loss function, denoted as \hingeone, aims to well rank peptides with significantly different binding 
affinities.
Given two peptides $\peptide_i$ and $\peptide_j$, this hinge loss function is defined as follows: 
\begin{equation}
\begin{aligned}
\label{eqn:hinge1}
\hingeone(\peptide_i,\peptide_j)= 
& \max(0, c+(\nbf_i-\nbf_j)-(\Sf(\peptide_i)-\Sf(\peptide_j )),
\text{ where }{\lev_i>\lev_j}, 
\end{aligned}
\end{equation}
where $\lev_i$ denotes the binding level of peptide $\peptide_i$ according to the Table~\ref{tbl:level}; $\lev_i > \lev_j$ denotes that 
the binding level of peptide $\peptide_i$ is higher than the peptide $\peptide_j$;
%\st{is a pre-specified constant to define the significance of the difference between binding affinities;} 
%
$\nbf_i$ and $\nbf_j$ are the ground-truth normalized binding affinities of $\peptide_i$ and $\peptide_j$, respectively;
$c>0$ is a pre-specified constant to increase the difference between two predicted scores. 
%
%\ziqi{\st{Normalized binding affinities $b$ range from 0.0 to 1.0, and thus the value of the difference $b_i-b_j$ can be extremely small for a pair of peptides
%which satisfy the requirement in }equation~\ref{eqn:hinge1}.}\xia{I do not know what this sentence means.... }
%%
%\ziqi{\st{In order to make the predicted scores in a pair to be significantly different, the pre-specified constant $c$ is necessary.}}
%\xia{I do not know what this sentence is for.... }
%
%\ziqi{In \hingeone, \st{only pairs of peptides with significant difference in binding affinities would contribute to the loss.}}
%
\hingeone learns from two peptides of different binding levels and defines a margin value between two peptides as the difference of their 
ground-truth binding affinities $\nbf_i-\nbf_j$ plus a constant $c$.
If two peptides $p_i$ and $p_j$ are on different binding levels $\lev_i > \lev_j$, and the difference of their predicted scores is smaller than the 
margin  $c + (\nbf_i-\nbf_j)$, 
this pair of peptides will contribute to the overall loss; otherwise, the loss of this pair will be 0.
Note that \hingeone is only defined on peptides of different binding levels. 
For the peptides with the same or similar binding affinities, {\hingeone} allows incorrect ranking among them.
%\xia{is it right? what constraint do you refer to? }
%\ziqi{I think it is right, I replaced "constraint" with "margin"}
%\xia{I am not able to understand what you mean here starting from line 188 
%so I cannot edit anything... 10/26 10:45am...}

%------------------------------------------------------------------------------------------------------
\paragraph{Level-based Hinge Loss Function}
%------------------------------------------------------------------------------------------------------
%
Instead of ranking with respect to the margin as in \hingeone, 
we relax the ranking criterion and use a margin according to the difference of binding levels (Table~\ref{tbl:level}). 
Thus, the second hinge loss, denoted as \hingetwo, is defined as follows:
\begin{equation}
\label{eqn:hinge2}
\hingetwo(\peptide_i,\peptide_j)= \max(0, r\times(\lev_i-\lev_j)- ( \Sf(\peptide_i)- \Sf(\peptide_j))), \text{where }\lev_i>\lev_j, 
\end{equation}
where $r>0$ is a constant. 
Given a pair of peptides in two different binding levels, similar to \hingeone, \hingetwo requires that if the difference of their predicted scores is
smaller than a margin, this pair of peptides will contribute to the overall loss; otherwise, the loss of these two peptides will be 0.
However, unlike \hingeone, the margin defined in \hingetwo depends on the difference of binding levels between two peptides (i.e., $r\times(\lev_i-\lev_j)$).
Therefore, in \hingetwo, the margin values of all the peptides $(\peptide_1, \peptide_2, \cdots, \peptide_n)$ on the level $\lev_i$ 
to any other peptides on the level $\lev_j$ will be the same (i.e., $r\times(\lev_i-\lev_j)$).
Note that \hingetwo is defined on peptides of different binding levels, and thus also allows incorrect ranking among peptides of same 
binding levels as in \hingeone; the difference with \hingeone is on how the margin is calculated. 
%

%------------------------------------------------------------------------------------------------------
\paragraph{Constrained Level-based Hinge Loss Function}
%------------------------------------------------------------------------------------------------------
%
The third hinge loss function \hingethree extends \hingetwo by adding a constraint that two peptides of a same binding level 
	can have similar predicted scores.
This hinge loss is defined as follows:
\begin{equation}
\label{eqn:hinge3}
\hingethree(\peptide_i,\peptide_j)= \left\{
\begin{aligned}
& \max(0, r\times(\lev_i-\lev_j)- (\Sf(\peptide_i)-\Sf(\peptide_j))), & \text{ if }\lev_i> \lev_j,  \\
& \max(0, |\Sf(\peptide_i)-\Sf(\peptide_j)|-r), & \text{ if } \lev_i=\lev_j. 
\end{aligned}
\right.
\end{equation}
%
%\xia{please double check all the loss functions 10/26 10:55am...}
%
%\ziqi{\st{We notice many noisy data when we preprocess the dataset such as conflicting entries with different binding 
%affinity entries}}\xia{simplify this sentence... }.
%\ziqi{\st{As we mentioned in Section, the IEDB dataset contains many conflicting entries.}}
%%
%\ziqi{\st{That makes us to consider that those binding affinity entries without confliction may also suffer such noise.}}
%\xia{right to the point; never use "make us to consider" -- it is not formal language... }
%
Given a pair of peptides on a same binding level, the added constraint (the case if $\lev_i = \lev_j$) requires that if the absolute difference 
$|\Sf(\peptide_i)-\Sf(\peptide_j)|$ is smaller than the pre-specified margin $r$, the loss will be zero; 
otherwise, this pair will have non-zero loss.
%
%\ziqi{\st{Therefore, we decide to merely limit the absolute value $|\Sf(\peptide_i)-\Sf(\peptide_j)|$ between pairs within the same level instead 
%of using the difference $|\Sf(\peptide_i)-\Sf(\peptide_j)|$ directly to well sort those pairs}
%\xia{I do not understand what you mean here... }}.
%
The constraint on the absolute difference allows incorrect ranking among peptides on a same binding 
level as long as their predicted scores are similar.
%

%====================================================================================
\subsubsection{Mean-Squares Loss}
\label{sec:methods:loss:meansquareloss}
%====================================================================================
%
We also compare a mean-squares loss function, denoted as \meansquare, proposed in~\cite{Donnell2018,Paul2019},
to fit the entries without exact binding affinity values as below: 
\begin{equation}
\label{eqn:meansquare}
\begin{aligned}
\meansquare(\peptide_i)= & \left\{
\begin{aligned}
& (\Sf(\peptide_i)-\nbf_i)^2 \quad && \text{if } \entry_i\ \text{is quantitative, } \\
& (\max(0,\Sf(\peptide_i)-\nbf_i))^2 \quad && \text{if } \entry_i\ \text{is qualitative and } l_i = 1\text{ (i.e., negative binding)},\\
& (\max(0,\nbf_i-\Sf(\peptide_i)))^2 \quad && \text{if } \entry_i\ \text{is qualitative and } l_i > 1\text{ (i.e., positive binding)}, \\
\end{aligned}
\right.
\end{aligned}
\end{equation}
where ``$\entry_i\ \text{is quantitative}$" denotes that the peptide-MHC entry $\entry_i$ is associated with an exact binding affinity value $x_i$.
In this case, the \meansquare loss is calculated as the squared difference between the predicted score $\Sf(\peptide_i)$ and $\nbf_i$ ($\nbf_i$ is 
normalized from $x_i$ as in Equation~\ref{eqn:clamp}). 
In Equation~\ref{eqn:meansquare}, 
``$\entry_i$ is qualitative" denotes that $\entry_i$ is associated with a binding level $\lev_i$ instead of a binding affinity value (Table~\ref{tbl:level}). 
In this case, $\nbf_i$ is normalized from the binding level thresholds (i.e., $x_i\in\{100, 500, 1000, 5000\}$ in calculating $\nbf_i$ in Equation~\ref{eqn:clamp}). 
When qualitative $\entry_i$ has $\lev_i = 1$, that is, the peptide does not bind to the allele and the binding affinity is low (i.e., large binding affinity value), 
the predicted score $\Sf(\peptide_i)$ should be small enough compared to $\nbf_i$ in order not to increase the loss. 
When quantitative $\entry_i$ has $\lev_i>1$, that is, the peptide binds to the allele with reasonably high binding affinity (i.e., small binding affinity value), 
the predicted score $\Sf(\peptide_i)$ should be large enough compared to $\nbf_i$ in order not to increase the loss. 
%
%$\nbf_i$ is the normalized binding affinity value of peptide $\peptide_i$ for quantitative entries, 
%%\xia{or binding level $l_i$ with values (100, 500, 1000, 5000) for qualitative entry. }
%\ziqi{or the normalized binding level threshold (100, 500, 1,000 or 5,000) for qualitative entries (i.e., $b_i=\text{clamp}(1-\log_{50,000}(x_i), 0, 1)$; 
%$x_i\in\{100, 500, 1000, 5000\}$ for qualitative entries).}
%%
%\ziqi{Since the qualitative entry $m_i$ on the level $l_i = 0$ has the binding affinity value $x_i > 5000$, the predicted normalized binding affinity values $\Sf(\peptide_i)$ will contribute to the loss if $\Sf(\peptide_i)$ is greater than $b_i$.
%%
%For the qualitative entry $m_i$ on the level $l_i > 0$ (i.e., $x_i < \{100, 500, 1000, 5000\}$), $\Sf(\peptide_i)$ will contribute to the loss if $\Sf(\peptide_i)$ is smaller than $b_i$.
%.}
%%
%%
%%
%%\st{$\entry_i\ >\ x_i$ and $\entry_i\ <\ x_i$ denote that the peptide-MHC entry $\entry_i$ is a qualitative value
%%mapped into the inequality with a quantitative value}
%%as in table~\ref{tbl:level}. 
%\xia{the notations are very confusing. Please change....}
%\xia{you need to give the intuition behind the mean squares here... }
%\xia{\xia{I am not able to understand what you mean here starting from line 235 
%so I cannot edit anything... 10/26 ...}}
%%
%%

Note that in \meansquare, the predicted score $\Sf(\peptide)$ needs to be normalized into range $[0,1]$. This is because $\nbf$ is in 
range $[0, 1]$ (Equation~\ref{eqn:clamp}) so that $\Sf(\peptide)$ needs to be in the same range and thus neither $\Sf(\peptide)$ nor $\nbf$ 
will dominate the squared errors due to substantially large or small values. 
However, in the three hinge loss functions (Equation~\ref{eqn:hinge1}, \ref{eqn:hinge2} and \ref{eqn:hinge3}), the potential different 
range between $\Sf(\peptide)$ and $\nbf$ or $\lev$ could be accommodated by the constant $c$ (Equation~\ref{eqn:hinge1}) or $r$
(Equation~\ref{eqn:hinge2} and \ref{eqn:hinge2}), respectively. 
In \meansquare, we use sigmoid function to normalize $\Sf(\peptide)$. 

\section{Experimental Settings}
\label{sec:experiments}
%%%%%%%%%%%%%%%%%%%%%%%%%%%%%%%%%%%%%%%%%%%%%%%%%%%%

%=============================================================================
\subsection{Baseline methods}
\label{sec:experiments:baseline}
%=============================================================================

%*******************************************************************************************************************
\subsubsection{Encoding Methods}
\label{sec:experiments:baseline:encoding}
%*******************************************************************************************************************

%\xia{not sure if it is a good idea to move this section here... need to discuss...}

Encoding methods represent each amino acid with a vector. Popular encoding methods used by the previous works 
include \blosum encoding \cite{Donnell2018,Jurtz2017,Nielsen2016}, \onehot encoding \cite{Phloyphisut2019, Bhattacharya2017} 
and \wordtovec embedding method \cite{Vang2017}.
\blosum encoding utilizes the BLOSUM62 matrix~\cite{Henikoff1992}, 
which measures the evolutionary divergence information among amino acids. 
We use the $i$-th row of the \blosum matrix as the feature of $i$-th amino acid.
%\st{In the \blosum matrix $M\in \R^{20\times 20}$, the $i$-th row represents the features of $i$-th amino acid} \xia{this is not correct... notations...}.
%
\onehot encoding represents the $i$-th natural amino acid with an one-hot vector, in which 
all elements are `0' except the $i$-th position as `1'.
\wordtovec learns the embeddings of amino acids from their contexts in protein sequences or peptide sequences.
This embedding method requires learning on a large corpus of amino acid sequences, 
and is much more complicated than \onehot.
However, it is demonstrated~\cite{Phloyphisut2019} 
that \wordtovec embedding method is comparable to \onehot encoding method, and  
%
%Therefore, we don't take word2vec embedding method into consideration, and simply compare BLOSUM encoding, one-hot encoding and deep encoding. 
therefore, we use \blosum encoding and \onehot encoding, but not \wordtovec. 
Besides the above encoding methods, we also evaluate another deep encoding method, denoted as \deep, 
in which the encoding of each amino acid is learned during the training process. 
\deep encoding is not deterministic and is learned during the training process; 
the dimension of embedding vector needs to be specified as a predefined hyper-parameter.
We also combine different representations of amino acid generated by the above three encoding methods. These combinations include
$\blosum$+\onehot, $\blosum$+\deep, $\onehot$+\deep and $\blosum$+$\onehot$+$\deep$, where
``+" represents concatenation of the embeddings of amino acid from different encoding methods.
%\xia{denote the method by the time you describe them using macros... I haven't significantly edited this sections but 
%It should be substantially simplified... please work on it...}
%\ziqi{I already used macros to describe the methods}
%%
%%\st{would also be explored to demonstrate whether combinations of embedding methods can benefit the model.}
%%\xia{what do yo mean by "+" here and what does "Deep" represent?}
%\\

%*******************************************************************************************************************
\subsubsection{Baseline Method - Local Connected Neural Networks \mhcflurry}
\label{sec:experiments:baseline:mhcflurry}
%*******************************************************************************************************************

\mhcflurry \cite{Donnell2018} is a state-of-the-art deep model with locally-connected layers
for peptide binding prediction.    
In \mhcflurry, all peptides of length 8 to 15 are padded into length 15 by keeping the first and last four 
residues and inserting the padding elements in the middle (e.g., "GGFVPNMLSV" is padded to "GGFVXXPNXXXMLSV"). 
%
%%\st{Specifically, the padding elements are inserted between the middle residues and four residues at 
%%the N-end and C-end of peptides.
%%
%For example, the peptide "GGFVPNMLSV" is padded to "GGFVXXPNXXXMLSV", 
%in which "X" is the padding element.}
%\xia{this paragraph is still too lengthy, please simplify!}
%
The padded sequences are encoded into a feature matrix $\matr{E}\in \R^{15\times20}$ using \blosum encoding method.
%
%%
%\ziqi{\st{ encodes the padded peptide sequence into a feature matrix %$\matr{E}\in \R^{15\times20}$ using 
% encoding method.
%%
%Then, }
\mhcflurry employs locally-connected layers to extract local feature patterns for each $k$-mers in peptide sequences.
Unlike CNN using common filters across all $k$-mer residues in peptides, locally-connected layers apply local filters for each $k$-mer
to encode the position-specific features.
%\st{Unlike CNN using common filters across all $k$-mer residues in peptides,
%locally-connected layers use different filters across different $k$-mers to encode the position-specific features 
%for each $k$-mer residues into an embedding.} 
%
The encoded feature embeddings for all $k$-mers are then concatenated into a vector, and fed into the fully-connected layer for binding prediction.
%
%\xia{The potential issues with MHCflurry...;Ziqi: I cannot figure out any potential issues...}
%
To the best of our knowledge, \mhcflurry is one of the best neural network model for allele-specific peptide 
binding prediction problem. 
%

%************************************************************************************
\subsection{Batch Generation}
\label{sec:experiments:batch}
%************************************************************************************

For models with \meansquare as the loss function,  
we randomly sample a batch of peptides as the training batch. 
For models with the proposed pair-wise hinge loss functions (\hingeone, \hingetwo, \hingethree), to reduce 
computational costs, we construct pairs of peptides for each training batch from a sampled batch of peptides. % instead of all peptides.
%
%This pair construction method can benefit our model by 
%\ziqi{adding more margin constraints towards the predicted score of each peptide in the 
%training batch.}
%
Specifically, for \hingeone and \hingetwo, each pair consists of two peptides from different binding levels;
and for \hingethree, the constructed pairs can consist of two peptides from the same or different binding levels.
%

%************************************************************************************
\subsection{Model Training}
\label{sec:experiments:network}
%************************************************************************************

We use 5-fold cross validation to tune the hyper-parameters of all methods through a grid search approach. 
For each allele, we run the grid search algorithm to find the optimal hyper-parameters 
for the allele-specific model.
%\st{The best result that each method can achieve among all possible sets of hyper-parameters is reported.}
%\xia{this is WRONG!}
%
We apply stochastic gradient descent (SGD) to optimize the loss functions.
We use 10\% of the training set as a validation set. This validation set is applied to adjust the learning rate dynamically 
and determine the early stopping of training process.
We set the dimension of \deep encoding method as 20, which is equal to the dimension of \blosum and \deep encoding method.
We also set both the constant $c$ in \hingeone and the constant $r$ in \hingetwo and \hingethree as $0.2$.
%

%************************************************************************************
\subsection{Evaluation Metrics}
\label{sec:experiments:metrics}
%************************************************************************************

We use 4 types of evaluation metrics, including average rank (\avgrank), hit rate (\hr), 
area under the roc curve (\auc), and \roc, to evaluate the performance of the various model 
architectures and loss functions.
Both \avgrank and \hr metrics are employed to measure the effectiveness of our model on the prioritization of promising 
peptides. 
Specifically, \avgrank metric measures the average overall rankings of promising peptides; 
\hr metric measures the ratio of promising peptides ranked at top; 
\auc metric measures the possibility that positive peptides are ranked higher than negative peptides;
\roc metric measures the ratio of positive peptides that are prioritized higher than top-$n$ false positive peptides. 
%\xia{you need to explain why ranking based metrics are necessary and AUC and ROC are not sufficient... }
%
%7 evaluation metrics, including 4 ranking-based metrics, area under the roc curve (AUC), 
%ROC5 and ROC10, to evaluate the performance of the various model architectures and loss functions.
%%
%These 4 ranking-based metrics include:
%1) average rank of peptides with binding affinity less than 100 (AR100); 
%2) hit rate of peptides with binding affinity less than 100 (HR100);
%3) average rank of peptides with binding affinity less than 500 (AR500);
%4) hit rate of peptides with binding affinity less than 500 (HR500).
%
%
We denote $\ranki$ as the rank of peptide $\peptidei$ based on their predicted scores, 
$\pset_{{h}}$ as the set of peptides with binding affinities 
smaller than \thd (e.g., $\thd = 500 \text{nM}$). Then $\avgrank_{\scriptsize{\thd}}$ (e.g., $\avgrank_{\scriptsize{500}}$) is defined as follows,
%\xia{set should not be bold but in $\mathcal{P}$ style... using $\scriptsize{\thd}$ when in subscripts...need to work on the macro define... please check the 
%compound optimization paper and see how the macros are defined... }
%\ziqi{macro fixed}
%
\begin{equation}
\label{eqn:avgrankx}
\avgrank_{h} = \frac{\sum_{{p_i \in \mathcal{P}_{{h}}}}\ranki}{|\mathcal{P}_{{h}}|}, \text{ where } \pset_{{h}} = 
\{\peptide_i |\ \forall \nbf_i < \thd\}, 
\end{equation}
where $|\pset_{{h}}|$ is the size of $\pset_{{h}}$. Smaller values of $\avgrank_{\scriptsize{\thd}}$ indicate that 
promising peptides are ranked higher in general, and thus better model performance. 
The hit rate $\hr_{h}$ (e.g., \hrfive) is defined as follows, 
\begin{equation}
\label{eqn:hr100}
\hr_{h} = \frac{|\pset_{t} \cap \pset_{h}|}{|\pset_{h}|}, \text{ where } \nthd=|\pset_{\scriptsize{h}}|
\end{equation}
where $\pset_{t}$ denotes the set of peptides with predicted scores ranked at top \nthd.  
Larger values of $\hr_{h}$ indicate that more promising peptides are prioritized to top-\nthd by the model, 
and thus better performance.
%
%
%\st{Both avgrank metrics and hr metrics assess the performance of models on peptides with positive-high or positive binding affinities}\xia{what do you mean? }.
%
%\xia{please fix all the subscripts... 11/22}

We use $\thd=500\text{nM}$ as the threshold to distinguish positive peptides and negative peptides, and apply two metrics for classification to 
evaluate the model performance. The first classification metric \auc is 
calculated as below,
\begin{equation}
\label{eqn:auc}
\auc=\frac{1}{|\pset_{\scriptsize{500}}|(|\pset|-|\pset_{\scriptsize{500}}|)}\sum_{i=1}^{\tiny{|\pset_{\scriptsize{500}}|}}
\sum_{j=1}^{\tiny{|\pset|-|\pset_{\scriptsize{500}}|}}\mathbbm{1}(\Sf(\peptide_i)>\Sf(\peptide_j)),
\end{equation}  
%\xia{fix all subscripts!} \ziqi{I used tiny to change the script size}
%
where \pset is the set of all peptides, and $|\pset|$ is the number of peptides in the dataset;
$\pset_{\scriptsize{500}}$ is the set of all positive peptides, and $|\pset_{\scriptsize{500}}|$ is the number of positive peptides; 
$\mathbbm{1}(\cdot)$ is an indicator function ($\mathbbm{1}(x) = 1$ if $x$ is true, otherwise 0). 
Larger values of \auc indicate that positive peptides are more likely to be ranked higher than negative peptides.
$\roc_{t}$ (e.g., \rocfive) score is the area under the roc curve up to $t$ false positives.
$\roc_{t}$ is calculated as below. 
%\xia{$n$ is abused... } \ziqi{I replace $n$ with $t$...}
%
\begin{equation}
\label{eqn:rocx}
\roc_{t}=\frac{1}{|\pset_{500}|\nthd}\sum_{i=1}^{\tiny{|\pset_{500}|}}
\sum_{j=1}^{t}\mathbbm{1}(\Sf(\peptide_i)>\Sf(\peptide_j))
\end{equation}
Larger values of $\roc_{\scriptsize{\nthd}}$ indicate that the model can prioritize more positive peptides up to first \nthd false positive peptides.
%
%Both \auc and $\roc_{\scriptsize{\nthd}}$ evaluate the model performance by the probability that positive peptides are ranked higher than negative peptides. 
%
%However, \auc calculates the probability among all positive peptides and negative peptides, while $\roc_{\scriptsize{\nthd}}$ focus on the probability among all positive peptides 
%and top n false positive peptides, and thus indicates the early retrieval performance. }
%
We use 7 metrics constructed from the above 4 types of metrics to evaluate the model performance. These 7 metrics include 
\avgrankone, \hrone, \avgrankfive, \hrfive, \auc, \rocfive and \rocten.

%%************************************************************************************
%\subsubsection{Hybrid Evaluation Metrics}
%\label{sec:experiments:combine}
%%************************************************************************************

%We tune all the hyperparameters in section~\ref{sec:experiments:network} for all 12 methods. 
%
%The evaluation metrics as in Section~\ref{sec:experiments:metrics} measure different 
%aspects of the model performance. 
In order to compare the models with respect to one single metric in a holistic way, 
we define a hybrid metric \hybrid by combining all the evaluation metrics . 
Given a model trained with a set of hyper-parameters $\mathcal{Y}$, 
we denote its performance on metric ``\metric'' (\metric=$\avgrank_h$, $\hr_h$, $\auc_{h}$, $\roc_h$) 
as $\metric(\mathcal{Y})$,
and the best metric value as $\text{best}_{\mathcal{Y}}(\metric)=\max_{\mathcal{Y}}(\metric(\mathcal{Y}))$.
Then, the hybrid metric \hybrid for a model with hyper-parameters $\mathcal{Y}$ is defined as below,
\begin{equation}
\label{eqn:para}
\hybrid(\mathcal{Y}) = \sum_{\metric}{\frac{\mathbb{I}(\downarrow\metric) \times (\metric(\mathcal{Y})-\text{best}_{\mathcal{Y}}(\metric))}{\text{best}_{\mathcal{Y}}(\metric)}}, 
\end{equation}
where $\mathbb{I}(\downarrow\metric)$ is an identity function: 
%indicates whether smaller values on metric $\metric$ indicate better performance:
$\mathbb{I}(\downarrow\metric) = +1$ if smaller values on metric $\metric$ indicate better performance; 
$\mathbb{I}(\downarrow\metric) = -1$ otherwise. 
For metrics \avgrankone and \avgrankfive, a smaller value represents a better model performance. 
%
%Therefore, \ziqi{$\text{sign}$ for these two AR metrics will be -1; for all other metrics which larger values represents better performance, $\text{sign}$ will be +1. }

%%%%%%%%%%%%%%%%%%%%%%%%%%%%%%%%%%%%%%%%%%%%%%%%%%%%%%%%%%%%%%%%%%%%%%%%%%%%%%%%%%%%%
\section{Experimental Results}
\label{sec:results}
%%%%%%%%%%%%%%%%%%%%%%%%%%%%%%%%%%%%%%%%%%%%%%%%%%%%%%%%%%%%%%%%%%%%%%%%%%%%%%%%%%%%%

We present the experimental results in this section. All the parameters used in the experiments are reported in the Appendix.
%

%************************************************************************************
\subsection{Model Architecture Comparison}
\label{sec:results:model}
%************************************************************************************

%%--------------------------------------------------------------------------------------------------------
%\subsubsection{Comparing Model Architectures}
%\label{sec:results:model:arch}
%%--------------------------------------------------------------------------------------------------------

We evaluate all the 12 possible combinations of the 3 model architectures (\convmodel, \spannyconvmodel,
\mhcflurry) and the 4 loss functions (\hingeone, \hingetwo, \hingethree, \meansquare) with all the encoding methods through 
5-fold cross validation. Table~\ref{tbl:perform:overall} presents the overall performance comparison with
$\blosum$+$\onehot$+$\deep$ encoding method (encoding method comparison will be presented later in Section~\ref{sec:results:embed}). 
%
%with \blosum encoding through 5-fold cross validation.
We apply the grid search to determine the optimal hyperparameters of each method on each allele
with respect to the hybrid metric \hybrid (Appendix Section~\ref{sec:appendix:parameter}), 
%
%
%\st{We report the results of methods with the optimal hyper-parameters selected by the highest values on the hybrid metric and other 4 metrics ($\avgrank_{100}$, $\hr_{100}$, \auc, $\roc_5$) across all 128 alleles.}
%\xia{do you mean this: we report the best performance of these methods in terms of the hybrid metric and all the four metrics 
%across all the 128 alleles. How did you select the hyper-parameters exactly? How do you use 5 different metrics? }
and report the best performance in Table~\ref{tbl:perform:overall}. 
% of these methods with the optimal hyperparameters in terms of the hybrid metric across all the 128 alleles.
%
%\st{Note that we do not report the average value of those metrics on all alleles, because the values of different alleles vary a lot.}
%\xia{what do you mean "average value of those metrics"?}
%Specifically, for each allele, we evaluate the performance of all methods on that allele and report the result of the model 
%of the best performance with respect to \hybrid. 
%Therefore, different alleles may have different optimal hyperparameters. }
%
%Note that we do not average the values of these metrics across all alleles, because the values on different alleles vary a lot.
%
We use \mhcflurry with \meansquare loss in \cite{Donnell2018} as the baseline, and calculate the  
percentage improvement of our methods over the baseline across 128 alleles.
%
%Table~\ref{tbl:perform:overall} presents the comparison. 
%
In Table~\ref{tbl:perform:overall}, the best model for each allele is selected with respect to \hybrid; the model performance is further 
evaluated using the 7 evaluation metrics. 
%present the average improvement percentages of our models 
%with hyper-parameters selected by the hybrid metric and a metrics over the baseline.
%

%\input{tables/results}
%\input{tables/blosum_onehot_learned_results}
%\input{blosum_onehot_learned_results}
\begin{table}[!h]
	%  \vspace{-5pt}    
	\centering
	\caption{Overall Performance Comparison (\hybrid; $\blosum$+$\onehot$+$\deep$)}
	\label{tbl:perform:overall}
	%\vspace{-10pt}
	\begin{small}
		\begin{threeparttable}
			\begin{tabular}{
					@{\hspace{3pt}}l@{\hspace{6pt}}
					@{\hspace{3pt}}l@{\hspace{6pt}}
					@{\hspace{3pt}}r@{\hspace{6pt}}
					@{\hspace{3pt}}r@{\hspace{6pt}}
					@{\hspace{3pt}}r@{\hspace{6pt}}
					@{\hspace{3pt}}r@{\hspace{6pt}}
					@{\hspace{3pt}}r@{\hspace{6pt}}
					@{\hspace{3pt}}r@{\hspace{6pt}}
					@{\hspace{3pt}}r@{\hspace{6pt}}
				}
				\toprule
				Model & loss & {$\avgrank_{100}$}  & {$\hr_{100}$} & {$\avgrank_{500}$} & {$\hr_{500}$} 
				& {\auc} & {$\roc_5$} & {$\roc_{10}$} \\        
				\midrule        
				\multirow{4}{*}{\convmodel}
				& \hingeone &   7.93 & 4.71 & 2.80 & 5.48 & 5.13 & 8.43 & 7.26\\
				& \hingetwo &   5.63 & 5.47 & 1.66 & 3.59 & 4.56 & 7.11 & 4.65\\
				& \hingethree & 6.35 & 5.70 & 0.99 & 2.59 & 4.16 & 4.69 & 4.42\\
				& \meansquare & -6.26 & 0.02 & -7.87 & -3.98 & 0.16 & -3.34 & -3.94\\
				
				%\cmidrule(lr){2-9}
				\midrule
				\multirow{4}{*}{\spannyconvmodel} 
				& \hingeone &   \textbf{11.58} & \textbf{10.47} & \textbf{7.28} & \textbf{8.28} & \textbf{6.70} & \textbf{17.10} & \textbf{14.42}\\
				& \hingetwo &   8.97 & 8.64 & 6.57 & 7.36 & 6.04 & 12.89 & 10.85\\
				& \hingethree & 10.01 & 8.87 & 4.73 & 6.00 & 6.00 & 14.01 & 11.36\\
				& \meansquare & 8.66 & 8.14 & 2.77 & 4.28 & 3.93 & 13.54 & 9.68\\
				%	\cmidrule(lr){2-9}
				\midrule
				\multirow{4}{*}{\mhcflurry}
				& \hingeone &   11.06 & 8.93 & 5.60 & 5.20 & 4.42 & 11.10 & 9.51\\
				& \hingetwo &   9.45 & 5.77 & 5.09 & 4.43 & 4.72 & 8.05 & 6.95\\
				& \hingethree & 8.83 & 6.35 & 4.54 & 5.73 & 4.52 & 7.10 & 5.88\\
				& \meansquare & 0.00 & 0.00 & 0.00 & 0.00 & 0.00 & 0.00 & 0.00\\    
				\bottomrule
			\end{tabular}
			\begin{footnotesize}
				\begin{tablenotes}
					\item 
					\!\!The values in the table are percentage improvement compared with the baseline  \mhcflurry with \meansquare. 
					Models are trained using  $\blosum$+$\onehot$+$\deep$ encoding methods, and 
					selected with respect to \hybrid and evaluated using the 7 evaluation metrics. 
					The best improvement with respect to each metric is \textbf{bold}. 
					\par
				\end{tablenotes}
			\end{footnotesize}
		\end{threeparttable}
	\end{small}
	%  \vspace{-10pt}    
\end{table}
%\label{tbl:perform:overall}

%
%
Table~\ref{tbl:perform:overall} shows that as for the model architectures, on average, 
\spannyconvmodel achieves the best performance overall among all three model architectures
(e.g., \spannyconvmodel with \meansquare has 8.66\% improvement in $\avgrank_{100}$ 
and 8.14\% in $\hr_{100}$ over \mhcflurry with \meansquare).
Please note that when we calculate the improvement, we exclude 
alleles on which our models achieve more than 150\%
improvement (typically no more than 15 such alleles under different metrics). 
This is to remove potential bias due to a few alleles on which the improvement is extremely substantial.  
%
%\xia{can you present the performance of \mhcflurry with \meansquare, that is, the AR100 values, instead of 0's?}
%\ziqi{No...As I said, the values of these metrics vary a lot on different alleles. It doesn't make sense to average the values on such a broad range.}
%
\spannyconvmodel performs better than \convmodel on average.
\spannyconvmodel extends \convmodel with global kernels to extract global features from the entire peptide sequences.
The better performance of \spannyconvmodel than that of \convmodel 
% (e.g., 18.36\% improvement compared with 7.71\% in ROC5 with \hingeone loss)
%\ziqi{I understand that. But the improvement value over \convmodel doesn't appear in the above table. 
%I recalculate the improvement of SpConvM with \hingeone over ConvM with \hingeone and it should be 5.59\%}) 
indicates that global features could capture useful information from entire peptide sequences, which are typically short,
for binding prediction. 
In addition, \mhcflurry outperforms \convmodel on average. 
%(e.g, 12.09\% improvement compared with 7.71\% in ROC5 with \hingeone loss).
%
%\xia{remind what \mhcflurry does that \convmodel does not...}
%\ziqi{I think I have explained it later. 
%
The difference between \mhcflurry and \convmodel is that \mhcflurry learns position-specific features via position-specific kernels, 
and \convmodel learns local features via kernels that are common to all the locations.
As demonstrated in other studies~\cite{Donnell2018} that certain positions of peptides are more critical for their binding to alleles, 
the better performance of \mhcflurry over \convmodel could be attributed to its position-specific feature learning capablity. 
Moreover, since the peptide sequences are usually short (8-15 amino-acid long), it is very likely that these short sequences do not have 
strong local patterns, and thus \convmodel could not capture a lot of useful local information. 
In comparison with \mhcflurry, \spannyconvmodel integrates both local features via its \convmodel component and global features via 
global kernels. Such integration could enable \spannyconvmodel to capture global information as compensation to local features, and thus 
to improve model performance. 
%
%
%It demonstrates that the position-specific features are more \xia{suitable???} for the modeling of binding events 
%than the local features extracted by convolutional layers.
%%
%The better performance of global features in \spannyconvmodel and position-specific features in \mhcflurry over local features in \convmodel
%may be due to the fact that most peptides are short sequences with length of 8-15.
%%
%Compared with long sequences, short sequences are less likely to contain some local binding patterns 
%which can be extracted by the shared local kernels in convolutional layers;
%thus, \convmodel with only local features doesn't perform as well as \spannyconvmodel and \mhcflurry. 
%%
%However, both the local features and global features can help model the binding events. 
%%
%Therefore, \spannyconvmodel with two types of features can better model the binding events than \mhcflurry 
%with only position-specific features.\\

In addition to using the hybrid metric \hybrid to determine the optimal hyperparameters, we also apply another four metrics
 $\avgrank_{100}$, \hrone, \auc and $\roc_5$
to select the hyperparameters. %  such that the models have the best performance on each allele with respect to each of the metrics.
The results of the best models in terms of these four metrics are presented 
in Table~\ref{tbl:perform:avg} ~\ref{tbl:perform:hr100} ~\ref{tbl:perform:auc} and ~\ref{tbl:perform:roc} in the 
Appendix, respectively.
The results show the same trend as that in Table~\ref{tbl:perform:overall}, that is, on average, 
\spannyconvmodel outperforms \convmodel and \mhcflurry on all 7 metrics and \hingeone loss function is the best among all loss functions.
%\xia{it is unclear how you select the results in these tables...}
%\xia{please format the tables in a same way as Table 4; you need to have more discussion about these three tables... }
%\ziqi{I'm not sure what kinds of discussion should be added here, as I have explained the possible reasons of this trend.}

%\xia{Three tables here}
%

%\input{tables/avg100results}
%\input{tables/hr100results}
%\input{tables/aucresults}
%\label{tbl:performance:auc}
%\input{tables/rocresults}

%
%\begin{figure}[!t]
%\centering
%\includegraphics[width=0.6\textwidth]{plots/dist}
%\caption{Performance Distribution in Comparison with \mhcflurry and \meansquare}
%\label{fig:dist}
%\end{figure}

\begin{figure}[!h]
\centering
         \includegraphics[width=0.5\textwidth]{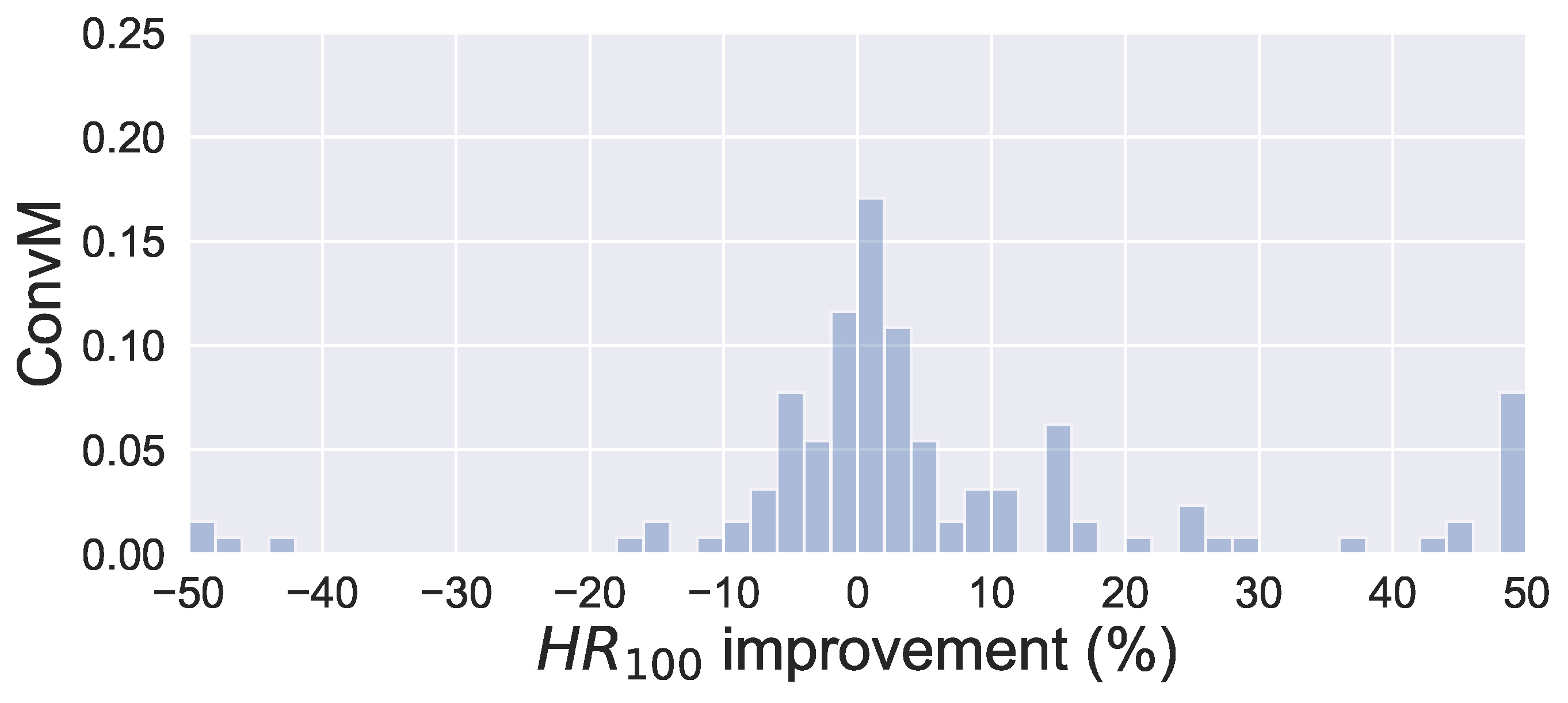}
         \caption{\convmodel Performance Improvement among All Alleles}
         \label{fig:dist:convm}
\end{figure}
\begin{figure}[!h]
\centering
         \includegraphics[width=0.5\textwidth]{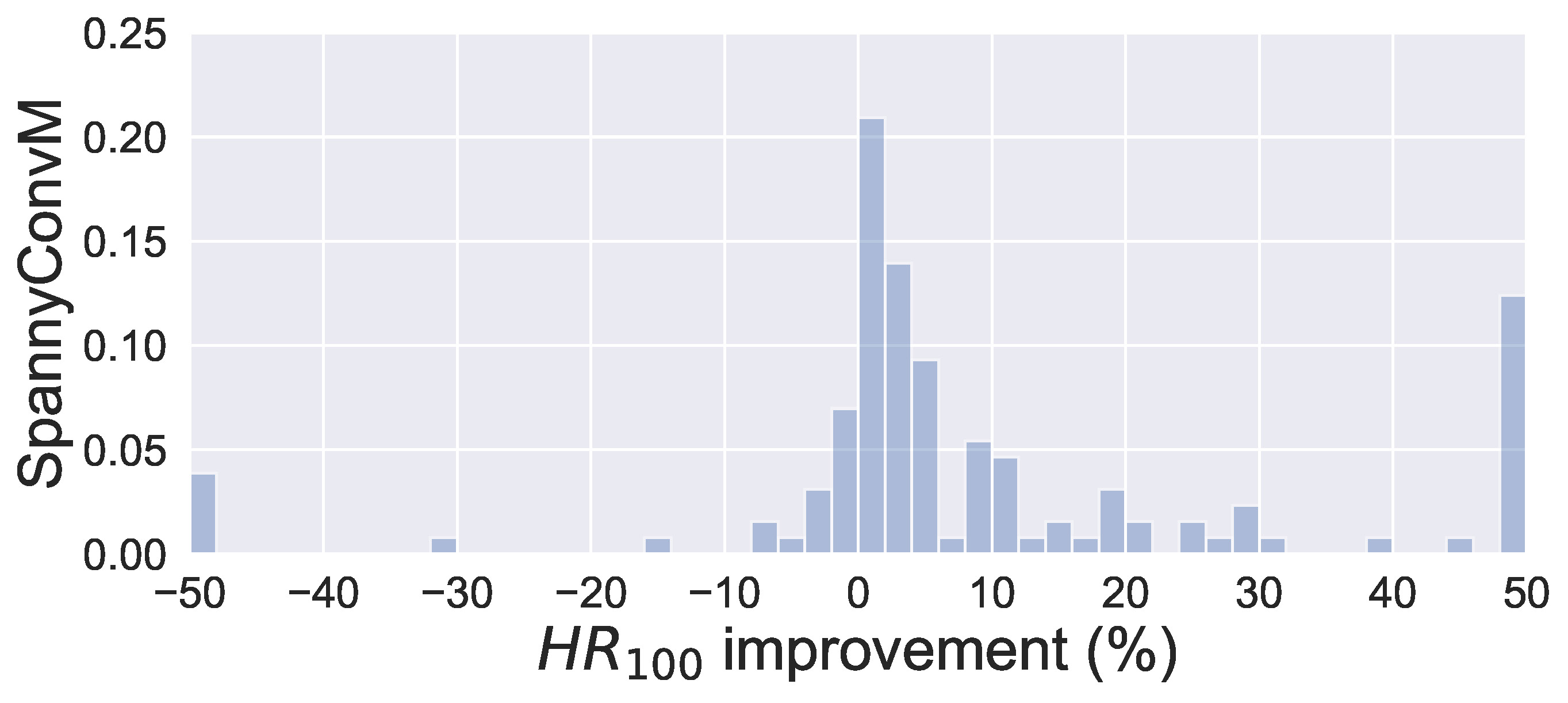}
         \caption{\spannyconvmodel Performance Improvement among All Alleles}
         \label{fig:dist:spanny}
\end{figure}
\begin{figure}[!h]
\centering
         \includegraphics[width=0.5\textwidth]{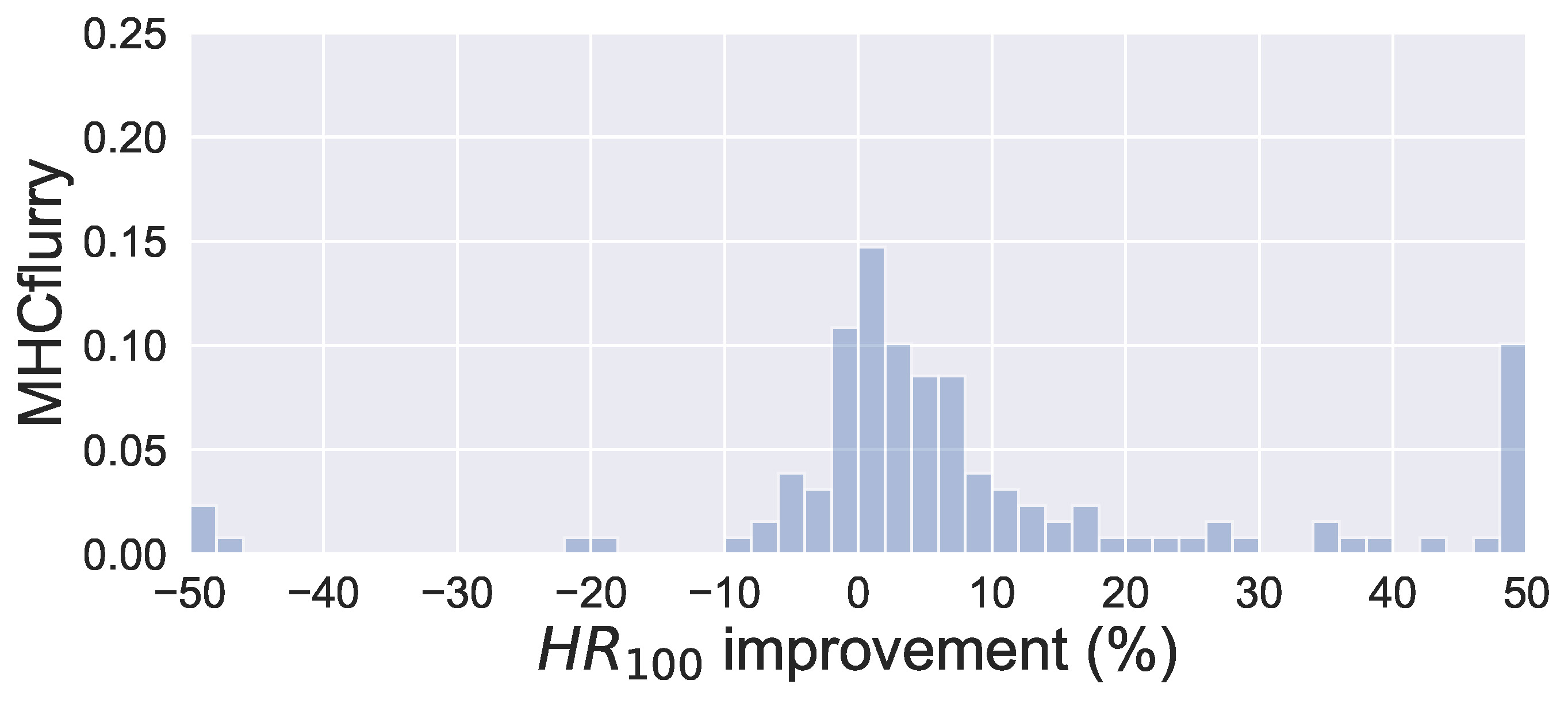}
         \caption{\mhcflurry Performance Improvement among All Alleles}
         \label{fig:dist:mhc}
\end{figure}

%\begin{figure}[!h]
%     \begin{subfigure}{0.3\textwidth}
%         \centering
%         \includegraphics[width=\textwidth]{plots/ConvM_HR100_dist.pdf}
%         \caption{\convmodel}
%         \label{fig:dist:convm}
%     \end{subfigure}
%     \hfill
%     \begin{subfigure}{0.3\textwidth}
%         \centering
%         \includegraphics[width=\textwidth]{SpannyConvM_HR100_dist.pdf}
%         \caption{\spannyconvmodel}
%         \label{fig:dist:spanny}
%     \end{subfigure}
%     \hfill
%     \begin{subfigure}{0.3\textwidth}
%         \centering
%         \includegraphics[width=\textwidth]{MHCflurry_HR100_dist.pdf}
%         \caption{\mhcflurry}
%         \label{fig:dist:mhc}
%     \end{subfigure}
%        \caption{Performance Improvement among All Alleles ($\blosum$+$\onehot$+$\deep$)}
%        \label{fig:dist}
%\end{figure}

Figure~\ref{fig:dist:convm}, \ref{fig:dist:spanny} and \ref{fig:dist:mhc} show the distributions of performance improvement among all the alleles from 
\convmodel, \spannyconvmodel and \mhcflurry with \hingeone, in comparison with \mhcflurry with \meansquare, respectively. 
All the methods use $\blosum$+$\onehot$+$\deep$ as encoding methods, and the performance 
is evaluated using $\hr_{100}$ in a same way as that in Table~\ref{tbl:perform:overall}. 
Figure~\ref{fig:dist:convm} shows that in \convmodel, about half of the alleles have performance improvement compared to 
that in \mhcflurry with \meansquare. Overall, there is an average 4.71\% improvement among all the alleles. 
Figure~\ref{fig:dist:spanny} shows that more alleles have performance improvement in \spannyconvmodel compared to that in \convmodel and 
in \mhcflurry with \meansquare, and more alleles have significant improvement. This indicates the strong performance of 
\spannyconvmodel. 
Figure~\ref{fig:dist:mhc} shows that in comparison with \meansquare as the loss function, \mhcflurry has more improvement using \hingeone
as the loss function (average improvement 8.93\%).

%************************************************************************************
\subsection{Loss Function Comparison}
\label{sec:results:loss}
%************************************************************************************

%
Table~\ref{tbl:perform:overall} also demonstrates that 
\hingeone is the most effective loss function in combination with each of the learning architectures, and 
all three hinge loss functions \hingeone, \hingetwo and \hingethree can outperform the \meansquare loss function. 
For example, for \spannyconvmodel, the performance improvement 
of \hingeone, \hingetwo, \hingethree and \meansquare in terms of $\avgrank_{100}$ follows the order  
$\hingeone (11.58\%) > \hingethree (10.01\%) > \hingetwo (8.97\%) > \meansquare (8.66\%)$, compared with the baseline \mhcflurry with \meansquare. 
This trend is also consistent for \convmodel and \mhcflurry. 
The better performance of \hingeone may be due to the use of a margin in the loss function that is determined by the 
binding affinity values (Equation~\ref{eqn:hinge1}).
This value-based margin could enforce granular ranking among peptides even when they are from a same binding level. 
In \hingetwo (Equation~\ref{eqn:hinge2}) and \hingethree (Equation~\ref{eqn:hinge3}), 
the margins are determined based on the levels of the binding affinities. While \hingetwo and \hingethree can still produce 
ranking structures of peptides according to their binding levels, they may fall short to differentiate peptides of a same binding level.

All the three hinge loss functions \hingeone, \hingetwo and \hingethree  outperform \meansquare across all the model architectures. 
This might be due to two reasons. 
%(1) 
First, the pairwise hinge loss functions are less sensitive to the imbalance of different amounts of peptides, 
either strongly binding or weakly/non-binding, by sampling and constructing pairs from respective peptides. 
Thus, the learning is not biased by one type of peptides, and the models can better 
learn the difference among different types of peptides, and accordingly produce better ranking orders of peptides.  
%(2)
Second, the pairwise hinge loss functions can tolerate insignificant measurement errors to some extent.
All the three hinge loss functions do not consider pairs of peptides with similar binding affinities.  
This enables our models to be more robust and tolerant to noisy data due to the measurement inaccuracy of binding affinities.
%
%

%************************************************************************************
\subsection{Encoding Method Comparison}
\label{sec:results:embed}
%************************************************************************************

We evaluate three encoding methods (\blosum, \onehot, \deep) and their combinations ($\blosum$+\deep, $\onehot$+\deep, $\blosum$+\onehot, 
$\blosum$+$\onehot$+$\deep$) over 
\spannyconvmodel with \hingeone loss (the best loss function overall) using \blosum through 5-fold cross validation.
We report the results of best models across all 128 alleles in the same way as in Section~\ref{sec:results:model}
(i.e., model selection with respect to \hybrid, evaluated using the 7 metrics).
%
%\xia{does \mhcflurry use \blosum? }
%We select BLOSUM encoding method as the baseline \ziqi{since most previous works use it to encode the peptide sequence}. \xia{why... }
%
Table~\ref{tbl:perform:encode_all} presents the average percentage improvement of the 7 encoding methods over the baseline on the 7 metrics.
The reported results in Table~\ref{tbl:perform:encode_all} are from the models with the optimal hyperparameters 
that are selected according to the hybrid metric \hybrid.
Table~\ref{tbl:perform:encode_all} shows that $\blosum$+$\onehot$+$\deep$ encoding method achieves the best performance in general.
$\blosum$+$\onehot$+$\deep$ encodes the amino acids using their inherent evolutionary information via \blosum
and identity of different amino acids via \onehot, both of which are deterministic and not specific to the learning problem,
and also the allele-specific information via \deep, which is learned in the model and thus specific to the learning problem. 
The combination of deterministic, amino acid identities and learned features enables very rich information content in the embeddings, and 
could be the reason why it outperforms others. 
With a similar rationale, $\blosum$+$\deep$ achieves the second best performance in general. 
\blosum on its own outperforms \onehot and \deep, respectively, indicating \blosum is rich in representing amino acid information. 
Combing \blosum with \onehot and \deep, respectively, introduces notable improvement over \blosum alone, indicating that 
$\blosum$+$\onehot$ and $\blosum$+$\deep$ are able to represent complementary information rather than that in \blosum. 
\onehot on itself alone performs the worst primarily due to its very limited information content. 
Combing $\onehot$ with \deep improves from \onehot but does not perform well compared to \deep alone. This may be due to that 
\onehot (i.e., amino acid identity) information still plays a substantial role in $\onehot$+$\deep$ so \deep information does not supply 
sufficient additional information. 
%
%\deep on its own performs slightly worse than the baseline \blosum, indicating that \blosum 
%
%
%%
%\ziqi{Including \onehot together with \blosum and \deep could slightly improve the performance in general.}
%\st{Including {\onehot} together with {\blosum} and {\deep} could not introduce much additional information
%but unnecessarily increase the number of parameters, which could lead to overfitting.}
%%

%\input{tables/spanny_embed_results}
%\input{spanny_embed_results}
\begin{table}[!h]   
	\centering
	\caption{Encoding Performance Comparison on \spannyconvmodel with \hingeone using \blosum (\hybrid)}
	\label{tbl:perform:encode_all}
	\begin{small}
		\begin{threeparttable}
			\begin{tabular}{
					@{\hspace{3pt}}l@{\hspace{6pt}}
					@{\hspace{3pt}}r@{\hspace{6pt}}
					@{\hspace{3pt}}r@{\hspace{6pt}}
					@{\hspace{3pt}}r@{\hspace{6pt}}
					@{\hspace{3pt}}r@{\hspace{6pt}}
					@{\hspace{3pt}}r@{\hspace{6pt}}
					@{\hspace{3pt}}r@{\hspace{6pt}}
					@{\hspace{3pt}}r@{\hspace{6pt}}
					@{\hspace{3pt}}r@{\hspace{6pt}}
				}
				\toprule
				encoding & {$\avgrank_{100}$}  & {$\hr_{100}$} & {$\avgrank_{500}$} & {$\hr_{500}$} 
				& {\auc} & {$\roc_5$} & {$\roc_{10}$} \\       
				\midrule        
				\blosum &       0.00 & 0.00 & 0.00 & 0.00 & 0.00 & 0.00 & 0.00\\
				\onehot &       -6.92 & -4.38 & -4.97 & -2.37 & -0.73 & -5.91 & -4.63\\
				\deep &         -3.33 & 0.69 & -2.76 & -0.56 & -0.15 & -1.22 & -1.63\\
				$\blosum$+$\onehot$ &       -0.96 & 0.95 & 0.21 & 0.79 & 0.3 & 1.57 & 0.89\\
				$\blosum$+$\deep$ & \textbf{1.37} & \textbf{1.79} & 0.69 & 1.49 & 0.61 & 3.39 & 2.49\\
				$\onehot$+$\deep$ & -4.72 & -1.65 & -3.37 & -1.26 & -0.32 & -3.25 & -2.67\\
				$\blosum$+$\onehot$+$\deep$ & -0.36 & 0.11 & \textbf{1.17} & \textbf{2.12} & \textbf{0.69} & \textbf{4.58} & \textbf{3.46}\\
				\bottomrule
			\end{tabular}
			\begin{footnotesize}
				\begin{tablenotes}
					\item 
					\!\!The values in the table are percentage improvement compared with \spannyconvmodel with \hingeone using \blosum. 
					Models are selected with respect to \hybrid and evaluated using the 7 evaluation metrics. 
					The best improvement with respect to each metric is \textbf{bold}. 
					\par
				\end{tablenotes}
			\end{footnotesize}
		\end{threeparttable}
	\end{small}
\end{table}
%  \label{tbl:perform:encode_all}

%

We also select the optimal set of hyperparameters with respect to $\avgrank_{100}$, $\hr_{100}$, \auc and {$\roc_5$}, 
and report the corresponding results  in Table~\ref{tbl:perform:encode_auc}, \ref{tbl:perform:embed_hr100}, 
\ref{tbl:perform:embed_roc5} and \ref{tbl:perform:embed_ar100} in Appendix, respectively.
With different model selection metrics, the encoding methods have different performance. However, in general, 
$\blosum$+$\onehot$+$\deep$ achieves better performance than other encoding methods over all the metrics. 

\subsection{Attention Weights}
\label{sec:experiments:weights}
%************************************************************************************

%%====================================================================================
%\subsubsection{Attention}
%\label{sec:experiments:analysis:attention}
%%====================================================================================

\begin{figure}[htbp] 
\centering
%\vspace{-20pt}
%\includegraphics[width=0.6\textwidth]{plots/conv_mat.pdf}
\includegraphics[width=0.6\textwidth]{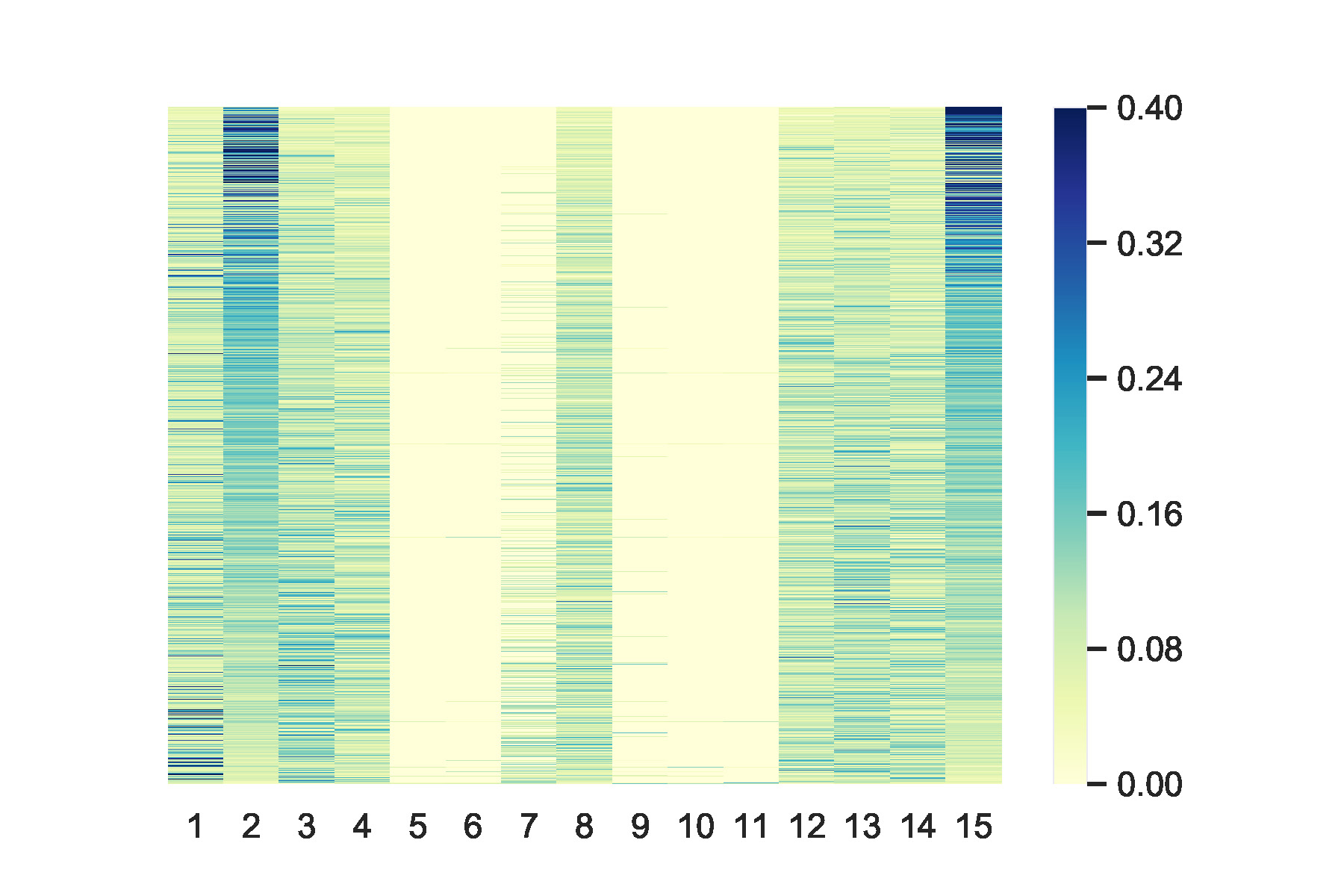}
\caption{Attention Weights for HLA-A-0201 Learned from \convmodel}
\label{fig:convattn}
\end{figure}

Figure~\ref{fig:convattn} shows the attention weight of HLA-A-0201 data learned by the attention 
layer of \convmodel (with $\blosum$+$\onehot$+$\deep$, \hingeone). 
In Figure~\ref{fig:convattn}, each column represents the weight of 1-mer embedding, that is, the embedding over one
amino acid, because the best kernel size for HLA-A-0201 data in \convmodel is 1; 
each row represents an attention weight learned for a specific peptide by \convmodel.
Figure~\ref{fig:convattn} shows the amino acids located at the second position and the last 
position contribute most to the binding events.
This is consistent with the conserved motif calculated by SMM matrix~\cite{Peters2005}. This indicates that \convmodel 
with the attention layer is able to accurately learn the importance of different positions in peptides in predicting peptide activities. 
% which is displayed in the 
%website of IEDB.
We do not present the attention weights learned by \spannyconvmodel, 
as in \spannyconvmodel, the attention weights do not show binding patterns as clear as those in \convmodel.
This is due to that \spannyconvmodel incorporates both local features and global features, and the global features 
might significantly contribute to the prediction and therefore the contribution from local features is reduced. 
\section{Conclusions}
\label{sec:discussion}
%%%%%%%%%%%%%%%%%%%%%%%%%%%%%%%%%%%%%%%%%%%%%%%%%%%%%%%%%%%%%%%%%%%%%%%%%%%%%%%%%%%%

Our methods contribute to the study of peptide-MHC binding prediction problem in two ways.
First, instead of predicting the exact binding affinities values as in the existing methods, we 
formulate the problem as to prioritize most possible peptide-MHC binding pairs via a ranking-based learning.  
We developed three pairwise ranking-based learning objectives for such prioritization, and the corresponding 
learning methods that impose the peptide-MHC pairs of higher binding affinities ranked above those with lower 
binding affinities with a certain margin. 
Our experimental results in comparison with the state-of-the-art regression based methods demonstrate the 
superior prediction performance of our methods in prioritizing and identifying the most likely binding peptides. 
In addition to the the learning objectives, we also developed two convolutional neural network-based model architectures 
\convmodel and \spannyconvmodel, which incorporate a new position encoding method and attention mechanism that 
differentiate the importance of amino acids at different positions in determining peptide-MHC binding. 
Our experiments show that the learned important positions and amino acids for allele HLA-A-0201 conform to the biological 
understanding of the allele. 
Our experimental results also demonstrate that our model architectures can achieve superior or at least comparable performance
with the state-of-the-art allele-specific baseline \mhcflurry.
%

%%%%%%%%%%%%%%%%%%%%%%%%%%%%%%%%%%%%%%%%%%%%%%%%%%%%%%%%%%%%%%%%%%%%%%%%%%%%%%%%%%%%
\section{Acknowledgments}
%%%%%%%%%%%%%%%%%%%%%%%%%%%%%%%%%%%%%%%%%%%%%%%%%%%%%%%%%%%%%%%%%%%%%%%%%%%%%%%%%%%%

This project was made possible, in part, by support from the National Science Foundation
under Grant Number IIS-1855501 and IIS-1827472, and the National Institute of General Medical Sciences
under Grant Number 2R01GM118470-05. 
Any opinions, findings, and conclusions or
recommendations expressed in this material are those of the authors and do not necessarily
reflect the views of the funding agencies.

\bibliographystyle{ACM-Reference-Format}
\bibliography{paper}


\begin{thebibliography}{31}

%%% the Web address.

\ifx \showCODEN    \undefined \def \showCODEN     #1{\unskip}     \fi
\ifx \showDOI      \undefined \def \showDOI       #1{#1}\fi
\ifx \showISBNx    \undefined \def \showISBNx     #1{\unskip}     \fi
\ifx \showISBNxiii \undefined \def \showISBNxiii  #1{\unskip}     \fi
\ifx \showISSN     \undefined \def \showISSN      #1{\unskip}     \fi
\ifx \showLCCN     \undefined \def \showLCCN      #1{\unskip}     \fi
\ifx \shownote     \undefined \def \shownote      #1{#1}          \fi
\ifx \showarticletitle \undefined \def \showarticletitle #1{#1}   \fi
\ifx \showURL      \undefined \def \showURL       {\relax}        \fi
% invisible to TeX
\providecommand\bibfield[2]{#2}
\providecommand\bibinfo[2]{#2}
\providecommand\natexlab[1]{#1}
\providecommand\showeprint[2][]{arXiv:#2}

\bibitem[\protect\citeauthoryear{Andreatta and Nielsen}{Andreatta and
  Nielsen}{2015}]%
        {Andreatta2015}
\bibfield{author}{\bibinfo{person}{Massimo Andreatta} {and}
  \bibinfo{person}{Morten Nielsen}.} \bibinfo{year}{2015}\natexlab{}.
\newblock \showarticletitle{Gapped sequence alignment using artificial neural
  networks: application to the {MHC} class I system}.
\newblock \bibinfo{journal}{\emph{Bioinformatics}} \bibinfo{volume}{32},
  \bibinfo{number}{4} (\bibinfo{date}{oct} \bibinfo{year}{2015}),
  \bibinfo{pages}{511--517}.
\newblock
\urldef\tempurl%
\url{https://doi.org/10.1093/bioinformatics/btv639}
\showDOI{\tempurl}


\bibitem[\protect\citeauthoryear{Bhattacharya, Sivakumar, Tokheim, Guthrie,
  Anagnostou, Velculescu, and Karchin}{Bhattacharya et~al\mbox{.}}{2017}]%
        {Bhattacharya2017}
\bibfield{author}{\bibinfo{person}{Rohit Bhattacharya}, \bibinfo{person}{Ashok
  Sivakumar}, \bibinfo{person}{Collin Tokheim}, \bibinfo{person}{Violeta~Beleva
  Guthrie}, \bibinfo{person}{Valsamo Anagnostou}, \bibinfo{person}{Victor~E.
  Velculescu}, {and} \bibinfo{person}{Rachel Karchin}.}
  \bibinfo{year}{2017}\natexlab{}.
\newblock \showarticletitle{Evaluation of machine learning methods to predict
  peptide binding to {MHC} Class I proteins}.
\newblock  (\bibinfo{date}{jun} \bibinfo{year}{2017}).
\newblock
\urldef\tempurl%
\url{https://doi.org/10.1101/154757}
\showDOI{\tempurl}


\bibitem[\protect\citeauthoryear{Blum, Wearsch, and Cresswell}{Blum
  et~al\mbox{.}}{2013}]%
        {Blum2013}
\bibfield{author}{\bibinfo{person}{Janice~S. Blum}, \bibinfo{person}{Pamela~A.
  Wearsch}, {and} \bibinfo{person}{Peter Cresswell}.}
  \bibinfo{year}{2013}\natexlab{}.
\newblock \showarticletitle{Pathways of Antigen Processing}.
\newblock \bibinfo{journal}{\emph{Annual Review of Immunology}}
  \bibinfo{volume}{31}, \bibinfo{number}{1} (\bibinfo{year}{2013}),
  \bibinfo{pages}{443--473}.
\newblock
\urldef\tempurl%
\url{https://doi.org/10.1146/annurev-immunol-032712-095910}
\showDOI{\tempurl}
\showeprint{https://doi.org/10.1146/annurev-immunol-032712-095910}
\newblock
\shownote{PMID: 23298205.}


\bibitem[\protect\citeauthoryear{Boehm, Bhinder, Raja, Dephoure, and
  Elemento}{Boehm et~al\mbox{.}}{2019}]%
        {Boehm2019}
\bibfield{author}{\bibinfo{person}{Kevin~Michael Boehm},
  \bibinfo{person}{Bhavneet Bhinder}, \bibinfo{person}{Vijay~Joseph Raja},
  \bibinfo{person}{Noah Dephoure}, {and} \bibinfo{person}{Olivier Elemento}.}
  \bibinfo{year}{2019}\natexlab{}.
\newblock \showarticletitle{Predicting peptide presentation by major
  histocompatibility complex class I: an improved machine learning approach to
  the immunopeptidome}.
\newblock \bibinfo{journal}{\emph{{BMC} Bioinformatics}} \bibinfo{volume}{20},
  \bibinfo{number}{1} (\bibinfo{date}{jan} \bibinfo{year}{2019}).
\newblock
\urldef\tempurl%
\url{https://doi.org/10.1186/s12859-018-2561-z}
\showDOI{\tempurl}


\bibitem[\protect\citeauthoryear{Bonsack, Hoppe, Winter, Tichy, Zeller,
  Küpper, Schitter, Blatnik, and Riemer}{Bonsack et~al\mbox{.}}{2019}]%
        {Bonsack2019}
\bibfield{author}{\bibinfo{person}{Maria Bonsack}, \bibinfo{person}{Stephanie
  Hoppe}, \bibinfo{person}{Jan Winter}, \bibinfo{person}{Diana Tichy},
  \bibinfo{person}{Christine Zeller}, \bibinfo{person}{Marius~D. Küpper},
  \bibinfo{person}{Eva~C. Schitter}, \bibinfo{person}{Renata Blatnik}, {and}
  \bibinfo{person}{Angelika~B. Riemer}.} \bibinfo{year}{2019}\natexlab{}.
\newblock \showarticletitle{Performance Evaluation of {MHC} Class-I Binding
  Prediction Tools Based on an Experimentally Validated
  {MHC}{\textendash}Peptide Binding Data Set}.
\newblock \bibinfo{journal}{\emph{Cancer Immunology Research}}
  \bibinfo{volume}{7}, \bibinfo{number}{5} (\bibinfo{date}{mar}
  \bibinfo{year}{2019}), \bibinfo{pages}{719--736}.
\newblock
\urldef\tempurl%
\url{https://doi.org/10.1158/2326-6066.cir-18-0584}
\showDOI{\tempurl}


\bibitem[\protect\citeauthoryear{Chorowski, Bahdanau, Serdyuk, Cho, and
  Bengio}{Chorowski et~al\mbox{.}}{2015}]%
        {Chorowski2015}
\bibfield{author}{\bibinfo{person}{Jan Chorowski}, \bibinfo{person}{Dzmitry
  Bahdanau}, \bibinfo{person}{Dmitriy Serdyuk}, \bibinfo{person}{Kyunghyun
  Cho}, {and} \bibinfo{person}{Yoshua Bengio}.}
  \bibinfo{year}{2015}\natexlab{}.
\newblock \showarticletitle{Attention-Based Models for Speech Recognition}.
\newblock  (\bibinfo{year}{2015}).
\newblock
\showeprint[arXiv]{http://arxiv.org/abs/1506.07503v1}~[cs.CL]


\bibitem[\protect\citeauthoryear{Couzin-Frankel}{Couzin-Frankel}{2013}]%
        {Couzin-Frankel1432}
\bibfield{author}{\bibinfo{person}{Jennifer Couzin-Frankel}.}
  \bibinfo{year}{2013}\natexlab{}.
\newblock \showarticletitle{Cancer Immunotherapy}.
\newblock \bibinfo{journal}{\emph{Science}} \bibinfo{volume}{342},
  \bibinfo{number}{6165} (\bibinfo{year}{2013}), \bibinfo{pages}{1432--1433}.
\newblock
\showISSN{0036-8075}
\urldef\tempurl%
\url{https://doi.org/10.1126/science.342.6165.1432}
\showDOI{\tempurl}
\showeprint{https://science.sciencemag.org/content/342/6165/1432.full.pdf}


\bibitem[\protect\citeauthoryear{Esfahani, Roudaia, Buhlaiga, {Del Rincon},
  Papneja, and {Miller Jr}}{Esfahani et~al\mbox{.}}{2020}]%
        {Esfahani2020}
\bibfield{author}{\bibinfo{person}{K Esfahani}, \bibinfo{person}{L Roudaia},
  \bibinfo{person}{N Buhlaiga}, \bibinfo{person}{S~V {Del Rincon}},
  \bibinfo{person}{N Papneja}, {and} \bibinfo{person}{W~H {Miller Jr}}.}
  \bibinfo{year}{2020}\natexlab{}.
\newblock \showarticletitle{{A review of cancer immunotherapy: from the past,
  to the present, to the future}}.
\newblock \bibinfo{journal}{\emph{Current oncology (Toronto, Ont.)}}
  \bibinfo{volume}{27}, \bibinfo{number}{Suppl 2} (\bibinfo{date}{apr}
  \bibinfo{year}{2020}), \bibinfo{pages}{S87--S97}.
\newblock
\showISSN{1718-7729}
\urldef\tempurl%
\url{https://doi.org/10.3747/co.27.5223}
\showDOI{\tempurl}


\bibitem[\protect\citeauthoryear{Goldberg and Levy}{Goldberg and Levy}{2014}]%
        {Goldberg2014}
\bibfield{author}{\bibinfo{person}{Yoav Goldberg} {and} \bibinfo{person}{Omer
  Levy}.} \bibinfo{year}{2014}\natexlab{}.
\newblock \showarticletitle{word2vec Explained: deriving Mikolov et al.'s
  negative-sampling word-embedding method}.
\newblock  (\bibinfo{year}{2014}).
\newblock
\showeprint[arXiv]{http://arxiv.org/abs/1402.3722v1}~[cs.CL]


\bibitem[\protect\citeauthoryear{Han and Kim}{Han and Kim}{2017}]%
        {Han2017}
\bibfield{author}{\bibinfo{person}{Youngmahn Han} {and}
  \bibinfo{person}{Dongsup Kim}.} \bibinfo{year}{2017}\natexlab{}.
\newblock \showarticletitle{Deep convolutional neural networks for pan-specific
  peptide-{MHC} class I binding prediction}.
\newblock \bibinfo{journal}{\emph{{BMC} Bioinformatics}} \bibinfo{volume}{18},
  \bibinfo{number}{1} (\bibinfo{date}{dec} \bibinfo{year}{2017}).
\newblock
\urldef\tempurl%
\url{https://doi.org/10.1186/s12859-017-1997-x}
\showDOI{\tempurl}


\bibitem[\protect\citeauthoryear{Henikoff and Henikoff}{Henikoff and
  Henikoff}{1992}]%
        {Henikoff1992}
\bibfield{author}{\bibinfo{person}{S. Henikoff} {and} \bibinfo{person}{J.~G.
  Henikoff}.} \bibinfo{year}{1992}\natexlab{}.
\newblock \showarticletitle{Amino acid substitution matrices from protein
  blocks.}
\newblock \bibinfo{journal}{\emph{Proceedings of the National Academy of
  Sciences}} \bibinfo{volume}{89}, \bibinfo{number}{22} (\bibinfo{date}{nov}
  \bibinfo{year}{1992}), \bibinfo{pages}{10915--10919}.
\newblock
\urldef\tempurl%
\url{https://doi.org/10.1073/pnas.89.22.10915}
\showDOI{\tempurl}


\bibitem[\protect\citeauthoryear{Hu, Wang, Hu, Wan, Chen, Xiong, Wang, Zhao,
  Huang, and Zeng}{Hu et~al\mbox{.}}{2018}]%
        {Hu2018}
\bibfield{author}{\bibinfo{person}{Yan Hu}, \bibinfo{person}{Ziqiang Wang},
  \bibinfo{person}{Hailin Hu}, \bibinfo{person}{Fangping Wan},
  \bibinfo{person}{Lin Chen}, \bibinfo{person}{Yuanpeng Xiong},
  \bibinfo{person}{Xiaoxia Wang}, \bibinfo{person}{Dan Zhao},
  \bibinfo{person}{Weiren Huang}, {and} \bibinfo{person}{Jianyang Zeng}.}
  \bibinfo{year}{2018}\natexlab{}.
\newblock \showarticletitle{{ACME}: Pan-specific peptide-{MHC} class I binding
  prediction through attention-based deep neural networks}.
\newblock  (\bibinfo{date}{nov} \bibinfo{year}{2018}).
\newblock
\urldef\tempurl%
\url{https://doi.org/10.1101/468363}
\showDOI{\tempurl}


\bibitem[\protect\citeauthoryear{Jurtz, Paul, Andreatta, Marcatili, Peters, and
  Nielsen}{Jurtz et~al\mbox{.}}{2017}]%
        {Jurtz2017}
\bibfield{author}{\bibinfo{person}{Vanessa Jurtz}, \bibinfo{person}{Sinu Paul},
  \bibinfo{person}{Massimo Andreatta}, \bibinfo{person}{Paolo Marcatili},
  \bibinfo{person}{Bjoern Peters}, {and} \bibinfo{person}{Morten Nielsen}.}
  \bibinfo{year}{2017}\natexlab{}.
\newblock \showarticletitle{{NetMHCpan}-4.0: Improved Peptide{\textendash}{MHC}
  Class I Interaction Predictions Integrating Eluted Ligand and Peptide Binding
  Affinity Data}.
\newblock \bibinfo{journal}{\emph{The Journal of Immunology}}
  \bibinfo{volume}{199}, \bibinfo{number}{9} (\bibinfo{date}{oct}
  \bibinfo{year}{2017}), \bibinfo{pages}{3360--3368}.
\newblock
\urldef\tempurl%
\url{https://doi.org/10.4049/jimmunol.1700893}
\showDOI{\tempurl}


\bibitem[\protect\citeauthoryear{Kim, Sidney, Buus, Sette, Nielsen, and
  Peters}{Kim et~al\mbox{.}}{2014}]%
        {Kim2014}
\bibfield{author}{\bibinfo{person}{Yohan Kim}, \bibinfo{person}{John Sidney},
  \bibinfo{person}{S{\o}ren Buus}, \bibinfo{person}{Alessandro Sette},
  \bibinfo{person}{Morten Nielsen}, {and} \bibinfo{person}{Bjoern Peters}.}
  \bibinfo{year}{2014}\natexlab{}.
\newblock \showarticletitle{Dataset size and composition impact the reliability
  of performance benchmarks for peptide-{MHC} binding predictions}.
\newblock \bibinfo{journal}{\emph{{BMC} Bioinformatics}} \bibinfo{volume}{15},
  \bibinfo{number}{1} (\bibinfo{year}{2014}), \bibinfo{pages}{241}.
\newblock
\urldef\tempurl%
\url{https://doi.org/10.1186/1471-2105-15-241}
\showDOI{\tempurl}


\bibitem[\protect\citeauthoryear{Kim, Sidney, Pinilla, Sette, and Peters}{Kim
  et~al\mbox{.}}{2009}]%
        {Kim2009}
\bibfield{author}{\bibinfo{person}{Yohan Kim}, \bibinfo{person}{John Sidney},
  \bibinfo{person}{Clemencia Pinilla}, \bibinfo{person}{Alessandro Sette},
  {and} \bibinfo{person}{Bjoern Peters}.} \bibinfo{year}{2009}\natexlab{}.
\newblock \showarticletitle{Derivation of an amino acid similarity matrix for
  peptide:{MHC} binding and its application as a Bayesian prior}.
\newblock \bibinfo{journal}{\emph{{BMC} Bioinformatics}} \bibinfo{volume}{10},
  \bibinfo{number}{1} (\bibinfo{year}{2009}), \bibinfo{pages}{394}.
\newblock
\urldef\tempurl%
\url{https://doi.org/10.1186/1471-2105-10-394}
\showDOI{\tempurl}


\bibitem[\protect\citeauthoryear{Kuksa, Min, Dugar, and Gerstein}{Kuksa
  et~al\mbox{.}}{2015}]%
        {Kuksa2015}
\bibfield{author}{\bibinfo{person}{Pavel~P. Kuksa},
  \bibinfo{person}{Martin~Renqiang Min}, \bibinfo{person}{Rishabh Dugar}, {and}
  \bibinfo{person}{Mark Gerstein}.} \bibinfo{year}{2015}\natexlab{}.
\newblock \showarticletitle{High-order neural networks and kernel methods for
  peptide-{MHC} binding prediction}.
\newblock \bibinfo{journal}{\emph{Bioinformatics}} (\bibinfo{date}{jul}
  \bibinfo{year}{2015}), \bibinfo{pages}{btv371}.
\newblock
\urldef\tempurl%
\url{https://doi.org/10.1093/bioinformatics/btv371}
\showDOI{\tempurl}


\bibitem[\protect\citeauthoryear{Lundegaard, Lamberth, Harndahl, Buus, Lund,
  and Nielsen}{Lundegaard et~al\mbox{.}}{2008}]%
        {Lundegaard2008}
\bibfield{author}{\bibinfo{person}{Claus Lundegaard}, \bibinfo{person}{Kasper
  Lamberth}, \bibinfo{person}{Mikkel Harndahl}, \bibinfo{person}{S{\o}ren
  Buus}, \bibinfo{person}{Ole Lund}, {and} \bibinfo{person}{Morten Nielsen}.}
  \bibinfo{year}{2008}\natexlab{}.
\newblock \showarticletitle{{NetMHC}-3.0: accurate web accessible predictions
  of human, mouse and monkey {MHC} class I affinities for peptides of length
  8{\textendash}11}.
\newblock \bibinfo{journal}{\emph{Nucleic Acids Research}}
  \bibinfo{volume}{36}, \bibinfo{number}{suppl{\_}2} (\bibinfo{date}{may}
  \bibinfo{year}{2008}), \bibinfo{pages}{W509--W512}.
\newblock
\urldef\tempurl%
\url{https://doi.org/10.1093/nar/gkn202}
\showDOI{\tempurl}


\bibitem[\protect\citeauthoryear{Mellman, Coukos, and Dranoff}{Mellman
  et~al\mbox{.}}{2011}]%
        {Mellman2011}
\bibfield{author}{\bibinfo{person}{Ira Mellman}, \bibinfo{person}{George
  Coukos}, {and} \bibinfo{person}{Glenn Dranoff}.}
  \bibinfo{year}{2011}\natexlab{}.
\newblock \showarticletitle{{Cancer immunotherapy comes of age}}.
\newblock \bibinfo{journal}{\emph{Nature}} \bibinfo{volume}{480},
  \bibinfo{number}{7378} (\bibinfo{year}{2011}), \bibinfo{pages}{480--489}.
\newblock
\showISSN{1476-4687}
\urldef\tempurl%
\url{https://doi.org/10.1038/nature10673}
\showDOI{\tempurl}


\bibitem[\protect\citeauthoryear{Nielsen and Andreatta}{Nielsen and
  Andreatta}{2016}]%
        {Nielsen2016}
\bibfield{author}{\bibinfo{person}{Morten Nielsen} {and}
  \bibinfo{person}{Massimo Andreatta}.} \bibinfo{year}{2016}\natexlab{}.
\newblock \showarticletitle{{NetMHCpan}-3.0: improved prediction of binding to
  {MHC} class I molecules integrating information from multiple receptor and
  peptide length datasets}.
\newblock \bibinfo{journal}{\emph{Genome Medicine}} \bibinfo{volume}{8},
  \bibinfo{number}{1} (\bibinfo{date}{mar} \bibinfo{year}{2016}).
\newblock
\urldef\tempurl%
\url{https://doi.org/10.1186/s13073-016-0288-x}
\showDOI{\tempurl}


\bibitem[\protect\citeauthoryear{O'Donnell, Rubinsteyn, Bonsack, Riemer,
  Laserson, and Hammerbacher}{O'Donnell et~al\mbox{.}}{2018}]%
        {Donnell2018}
\bibfield{author}{\bibinfo{person}{Timothy~J. O'Donnell}, \bibinfo{person}{Alex
  Rubinsteyn}, \bibinfo{person}{Maria Bonsack}, \bibinfo{person}{Angelika~B.
  Riemer}, \bibinfo{person}{Uri Laserson}, {and} \bibinfo{person}{Jeff
  Hammerbacher}.} \bibinfo{year}{2018}\natexlab{}.
\newblock \showarticletitle{{MHCflurry}: Open-Source Class I {MHC} Binding
  Affinity Prediction}.
\newblock \bibinfo{journal}{\emph{Cell Systems}} \bibinfo{volume}{7},
  \bibinfo{number}{1} (\bibinfo{date}{jul} \bibinfo{year}{2018}),
  \bibinfo{pages}{129--132.e4}.
\newblock
\urldef\tempurl%
\url{https://doi.org/10.1016/j.cels.2018.05.014}
\showDOI{\tempurl}


\bibitem[\protect\citeauthoryear{O’Donnell, Rubinsteyn, and
  Laserson}{O’Donnell et~al\mbox{.}}{2020}]%
        {Odonnell2020}
\bibfield{author}{\bibinfo{person}{Timothy~J. O’Donnell},
  \bibinfo{person}{Alex Rubinsteyn}, {and} \bibinfo{person}{Uri Laserson}.}
  \bibinfo{year}{2020}\natexlab{}.
\newblock \showarticletitle{MHCflurry 2.0: Improved Pan-Allele Prediction of
  MHC Class I-Presented Peptides by Incorporating Antigen Processing}.
\newblock \bibinfo{journal}{\emph{Cell Systems}} \bibinfo{volume}{11},
  \bibinfo{number}{1} (\bibinfo{year}{2020}), \bibinfo{pages}{42 -- 48.e7}.
\newblock
\showISSN{2405-4712}
\urldef\tempurl%
\url{https://doi.org/10.1016/j.cels.2020.06.010}
\showDOI{\tempurl}


\bibitem[\protect\citeauthoryear{Paul, Croft, Purcell, Tscharke, Sette,
  Nielsen, and Peters}{Paul et~al\mbox{.}}{2019}]%
        {Paul2019}
\bibfield{author}{\bibinfo{person}{Sinu Paul}, \bibinfo{person}{Nathan~P.
  Croft}, \bibinfo{person}{Anthony~W. Purcell}, \bibinfo{person}{David~C.
  Tscharke}, \bibinfo{person}{Alessandro Sette}, \bibinfo{person}{Morten
  Nielsen}, {and} \bibinfo{person}{Bjoern Peters}.}
  \bibinfo{year}{2019}\natexlab{}.
\newblock \showarticletitle{Benchmarking predictions of {MHC} class I
  restricted T cell epitopes}.
\newblock  (\bibinfo{date}{jul} \bibinfo{year}{2019}).
\newblock
\urldef\tempurl%
\url{https://doi.org/10.1101/694539}
\showDOI{\tempurl}


\bibitem[\protect\citeauthoryear{Peters and Sette}{Peters and Sette}{2005}]%
        {Peters2005}
\bibfield{author}{\bibinfo{person}{Bjoern Peters} {and}
  \bibinfo{person}{Alessandro Sette}.} \bibinfo{year}{2005}\natexlab{}.
\newblock \showarticletitle{Generating quantitative models describing the
  sequence specificity of biological processes with the stabilized matrix
  method}.
\newblock \bibinfo{journal}{\emph{{BMC} Bioinformatics}} \bibinfo{volume}{6},
  \bibinfo{number}{1} (\bibinfo{year}{2005}), \bibinfo{pages}{132}.
\newblock
\urldef\tempurl%
\url{https://doi.org/10.1186/1471-2105-6-132}
\showDOI{\tempurl}


\bibitem[\protect\citeauthoryear{Phloyphisut, Pornputtapong, Sriswasdi, and
  Chuangsuwanich}{Phloyphisut et~al\mbox{.}}{2019}]%
        {Phloyphisut2019}
\bibfield{author}{\bibinfo{person}{Poomarin Phloyphisut},
  \bibinfo{person}{Natapol Pornputtapong}, \bibinfo{person}{Sira Sriswasdi},
  {and} \bibinfo{person}{Ekapol Chuangsuwanich}.}
  \bibinfo{year}{2019}\natexlab{}.
\newblock \showarticletitle{{MHCSeqNet}: a deep neural network model for
  universal {MHC} binding prediction}.
\newblock \bibinfo{journal}{\emph{{BMC} Bioinformatics}} \bibinfo{volume}{20},
  \bibinfo{number}{1} (\bibinfo{date}{may} \bibinfo{year}{2019}).
\newblock
\urldef\tempurl%
\url{https://doi.org/10.1186/s12859-019-2892-4}
\showDOI{\tempurl}


\bibitem[\protect\citeauthoryear{Purcell, McCluskey, and Rossjohn}{Purcell
  et~al\mbox{.}}{2007}]%
        {Purcell2007}
\bibfield{author}{\bibinfo{person}{Anthony~W. Purcell}, \bibinfo{person}{James
  McCluskey}, {and} \bibinfo{person}{Jamie Rossjohn}.}
  \bibinfo{year}{2007}\natexlab{}.
\newblock \showarticletitle{More than one reason to rethink the use of peptides
  in vaccine design}.
\newblock \bibinfo{journal}{\emph{Nature Reviews Drug Discovery}}
  \bibinfo{volume}{6}, \bibinfo{number}{5} (\bibinfo{date}{may}
  \bibinfo{year}{2007}), \bibinfo{pages}{404--414}.
\newblock
\urldef\tempurl%
\url{https://doi.org/10.1038/nrd2224}
\showDOI{\tempurl}


\bibitem[\protect\citeauthoryear{Valitutti, M{\"u}ller, Cella, Padovan, and
  Lanzavecchia}{Valitutti et~al\mbox{.}}{1995}]%
        {valitutti1995serial}
\bibfield{author}{\bibinfo{person}{Salvatore Valitutti},
  \bibinfo{person}{Sabina M{\"u}ller}, \bibinfo{person}{Marina Cella},
  \bibinfo{person}{Elisabetta Padovan}, {and} \bibinfo{person}{Antonio
  Lanzavecchia}.} \bibinfo{year}{1995}\natexlab{}.
\newblock \showarticletitle{Serial triggering of many T-cell receptors by a few
  peptide--MHC complexes}.
\newblock \bibinfo{journal}{\emph{Nature}} \bibinfo{volume}{375},
  \bibinfo{number}{6527} (\bibinfo{year}{1995}), \bibinfo{pages}{148--151}.
\newblock


\bibitem[\protect\citeauthoryear{Vang and Xie}{Vang and Xie}{2017}]%
        {Vang2017}
\bibfield{author}{\bibinfo{person}{Yeeleng~S Vang} {and}
  \bibinfo{person}{Xiaohui Xie}.} \bibinfo{year}{2017}\natexlab{}.
\newblock \showarticletitle{{HLA} class I binding prediction via convolutional
  neural networks}.
\newblock \bibinfo{journal}{\emph{Bioinformatics}} \bibinfo{volume}{33},
  \bibinfo{number}{17} (\bibinfo{date}{apr} \bibinfo{year}{2017}),
  \bibinfo{pages}{2658--2665}.
\newblock
\urldef\tempurl%
\url{https://doi.org/10.1093/bioinformatics/btx264}
\showDOI{\tempurl}


\bibitem[\protect\citeauthoryear{Vita, Mahajan, Overton, Dhanda, Martini,
  Cantrell, Wheeler, Sette, and Peters}{Vita et~al\mbox{.}}{2018}]%
        {Vita2018}
\bibfield{author}{\bibinfo{person}{Randi Vita}, \bibinfo{person}{Swapnil
  Mahajan}, \bibinfo{person}{James~A Overton}, \bibinfo{person}{Sandeep~Kumar
  Dhanda}, \bibinfo{person}{Sheridan Martini}, \bibinfo{person}{Jason~R
  Cantrell}, \bibinfo{person}{Daniel~K Wheeler}, \bibinfo{person}{Alessandro
  Sette}, {and} \bibinfo{person}{Bjoern Peters}.}
  \bibinfo{year}{2018}\natexlab{}.
\newblock \showarticletitle{The Immune Epitope Database ({IEDB}): 2018 update}.
\newblock \bibinfo{journal}{\emph{Nucleic Acids Research}}
  \bibinfo{volume}{47}, \bibinfo{number}{D1} (\bibinfo{date}{oct}
  \bibinfo{year}{2018}), \bibinfo{pages}{D339--D343}.
\newblock
\urldef\tempurl%
\url{https://doi.org/10.1093/nar/gky1006}
\showDOI{\tempurl}


\bibitem[\protect\citeauthoryear{Waldman, Fritz, and Lenardo}{Waldman
  et~al\mbox{.}}{2020}]%
        {Waldman2020}
\bibfield{author}{\bibinfo{person}{Alex~D Waldman}, \bibinfo{person}{Jill~M
  Fritz}, {and} \bibinfo{person}{Michael~J Lenardo}.}
  \bibinfo{year}{2020}\natexlab{}.
\newblock \showarticletitle{{A guide to cancer immunotherapy: from T cell basic
  science to clinical practice}}.
\newblock \bibinfo{journal}{\emph{Nature Reviews Immunology}}
  \bibinfo{volume}{20}, \bibinfo{number}{11} (\bibinfo{year}{2020}),
  \bibinfo{pages}{651--668}.
\newblock
\showISSN{1474-1741}
\urldef\tempurl%
\url{https://doi.org/10.1038/s41577-020-0306-5}
\showDOI{\tempurl}


\bibitem[\protect\citeauthoryear{Zeng and Gifford}{Zeng and Gifford}{2019}]%
        {Zeng2019}
\bibfield{author}{\bibinfo{person}{Haoyang Zeng} {and} \bibinfo{person}{David~K
  Gifford}.} \bibinfo{year}{2019}\natexlab{}.
\newblock \showarticletitle{{DeepLigand}: accurate prediction of {MHC} class I
  ligands using peptide embedding}.
\newblock \bibinfo{journal}{\emph{Bioinformatics}} \bibinfo{volume}{35},
  \bibinfo{number}{14} (\bibinfo{date}{jul} \bibinfo{year}{2019}),
  \bibinfo{pages}{i278--i283}.
\newblock
\urldef\tempurl%
\url{https://doi.org/10.1093/bioinformatics/btz330}
\showDOI{\tempurl}


\bibitem[\protect\citeauthoryear{Zhao and Sher}{Zhao and Sher}{2018}]%
        {Zhao2018}
\bibfield{author}{\bibinfo{person}{Weilong Zhao} {and} \bibinfo{person}{Xinwei
  Sher}.} \bibinfo{year}{2018}\natexlab{}.
\newblock \showarticletitle{Systematically benchmarking peptide-{MHC} binding
  predictors: From synthetic to naturally processed epitopes}.
\newblock \bibinfo{journal}{\emph{{PLOS} Computational Biology}}
  \bibinfo{volume}{14}, \bibinfo{number}{11} (\bibinfo{date}{nov}
  \bibinfo{year}{2018}), \bibinfo{pages}{e1006457}.
\newblock
\urldef\tempurl%
\url{https://doi.org/10.1371/journal.pcbi.1006457}
\showDOI{\tempurl}


\end{thebibliography}

%%%%%%%%%%%%%%%%%%%%%%%%%%%%%%%%%%%%%%%%%%%%%%%%%
\begin{appendix}
%%%%%%%%%%%%%%%%%%%%%%%%%%%%%%%%%%%%%%%%%%%%%%%%%
\section*{Appendix}

\setcounter{table}{0}
\renewcommand{\thetable}{A\arabic{table}}

\setcounter{figure}{0}
\renewcommand{\thefigure}{A\arabic{figure}}

\setcounter{section}{0}
\renewcommand{\thesection}{A\arabic{section}}

%\setcounter{algorithm}{0}
%\renewcommand{\thealgorithm}{A\arabic{algorithm}}

%========================================================================
\section{Additional Experimental Results}
%========================================================================

%------------------------------------------------------------------------------------------------------------------------------
\subsection{Hyperparameter search space}
\label{sec:appendix:parameter}
%------------------------------------------------------------------------------------------------------------------------------

The search space of hyperparameters includes batch size \{32, 128\}, number of layers \{1, 2\}, filter size \{1, 3, 5\}, number of filters \{8, 16, 64\}. 

%------------------------------------------------------------------------------------------------------------------------------
\subsection{Model Architecture Comparison}
\label{sec:appendix:model}
%------------------------------------------------------------------------------------------------------------------------------
%
Table~\ref{tbl:perform:avg}, \ref{tbl:perform:hr100}, \ref{tbl:perform:auc} and \ref{tbl:perform:roc} present
the average percentage improvement of our models over the baseline \mhcflurry with \meansquare.
Models are trained with \blosum+\onehot+\deep and selected with respect to 4 different metrics (i.e., $\avgrank_{100}$, $\hr_{100}$, 
\auc and $\roc_{5}$, respectively).
%\xia{what? Please format the tables according to Table~\ref{tbl:perform:avg}. } 

%\input{tables/blosum_onehot_learned_avg100results}
%% \label{tbl:perform:avg}
%%
%\input{tables/blosum_onehot_learned_hr100results}
%%\label{tbl:perform:hr100} 
%%
%\input{tables/blosum_onehot_learned_aucresults}
%%\label{tbl:perform:auc} 
%%
%\input{tables/blosum_onehot_learned_rocresults}
%%label{tbl:perform:roc}

%------------------------------------------------------------------------------------------------------------------------------
\subsection{Encoding Method Comparison}
%------------------------------------------------------------------------------------------------------------------------------

Table~\ref{tbl:perform:embed_ar100}, \ref{tbl:perform:embed_hr100}, \ref{tbl:perform:encode_auc},  and \ref{tbl:perform:embed_roc5} present 
the average percentage improvement of the 7 encoding methods over the baseline encoding method \blosum.
We used \spannyconvmodel with \hingeone as the model to evaluate encoding methods and their combination. 
We trained the models and selected the hyper-parameters with respect to 4 different metrics (i.e., $\avgrank_{100}$, $\hr_{100}$, 
\auc and $\roc_{5}$, respectively).
%\xia{what? Please format the tables according to Table~\ref{tbl:perform:avg}. } 

%\input{tables/blosum_onehot_learned_avg100results}
\begin{table}[!h]
	\centering
	\caption{Performance Comparison over \mhcflurry with \meansquare  ($\avgrank_{100}$; $\blosum$+$\onehot$+$\deep$)}
	\label{tbl:perform:avg}
	\begin{small}
		\begin{threeparttable}
			\begin{tabular}{
					@{\hspace{3pt}}l@{\hspace{6pt}}
					@{\hspace{3pt}}l@{\hspace{6pt}}
					@{\hspace{3pt}}r@{\hspace{6pt}}
					@{\hspace{3pt}}r@{\hspace{6pt}}
					@{\hspace{3pt}}r@{\hspace{6pt}}
					@{\hspace{3pt}}r@{\hspace{6pt}}
					@{\hspace{3pt}}r@{\hspace{6pt}}
					@{\hspace{3pt}}r@{\hspace{6pt}}
					@{\hspace{3pt}}r@{\hspace{6pt}}
				}
				\toprule
				Model & loss & {$\avgrank_{100}$}  & {$\hr_{100}$} & {$\avgrank_{500}$} & {$\hr_{500}$} 
				& {\auc} & {$\roc_5$} & {$\roc_{10}$} \\ 
				\midrule        
				\multirow{4}{*}{\convmodel}
				& \hingeone &   8.08 & 4.79 & 3.56 & 5.93 & 6.36 & 9.92 & 10.01\\
				& \hingetwo &   6.36 & 4.50 & 1.33 & 4.50 & 5.21 & 6.57 & 5.92\\
				& \hingethree & 6.58 & 3.49 & 1.36 & 3.26 & 5.63 & 7.25 & 6.05\\
				& \meansquare & -5.17 & -7.93 & -6.69 & -1.93 & 1.24 & -3.51 & -3.36\\
				
				%\cmidrule(lr){2-9}
				\midrule
				\multirow{4}{*}{\spannyconvmodel} 
				& \hingeone &   10.75 & 7.58 & \textbf{8.06} & \textbf{9.61} & \textbf{8.68} & \textbf{18.84} & \textbf{16.29}\\
				& \hingetwo &   10.10 & 5.39 & 5.87 & 6.69 & 7.39 & 14.09 & 10.97\\
				& \hingethree & 10.59 & \textbf{8.28} & 5.32 & 7.47 & 7.35 & 14.66 & 11.56\\
				& \meansquare & 7.36 & 7.54 & 3.27 & 5.74 & 5.16 & 15.66 & 11.68\\
				%	\cmidrule(lr){2-9}
				\midrule
				\multirow{4}{*}{\mhcflurry}
				& \hingeone &   \textbf{10.95} & 6.74 & 6.62 & 7.95 & 6.45 & 15.87 & 11.45\\
				& \hingetwo &   8.61 & 7.32 & 5.37 & 6.08 & 5.71 & 10.75 & 9.39\\
				& \hingethree & 8.20 & 4.19 & 4.87 & 5.87 & 4.92 & 11.79 & 7.99\\
				& \meansquare & 0.00 & 0.00 & 0.00 & 0.00 & 0.00 & 0.00 & 0.00\\
				\bottomrule
			\end{tabular}
			\begin{footnotesize}
				\begin{tablenotes}
					\item 
					\!\!The values in the table are percentage improvement compared with the baseline  \mhcflurry with \meansquare. 
					Models are trained using  $\blosum$+$\onehot$+$\deep$ encoding methods, and 
					selected with respect to $\avgrank_{100}$ and evaluated using the 7 evaluation metrics. 
					The best improvement with respect to each metric is \textbf{bold}. 
					\par
				\end{tablenotes}
			\end{footnotesize}
		\end{threeparttable}
	\end{small}    
\end{table}
% \label{tbl:perform:avg}
%
%\input{tables/blosum_onehot_learned_hr100results}
\begin{table}[!h]
	%  \vspace{-5pt}    
	\centering
	\caption{Performance Comparison over \mhcflurry with \meansquare  ($\hr_{100}$; $\blosum$+$\onehot$+$\deep$)}
	\label{tbl:perform:hr100}
	%\vspace{-10pt}
	\begin{small}
		\begin{threeparttable}
			\begin{tabular}{
					@{\hspace{3pt}}l@{\hspace{6pt}}
					@{\hspace{3pt}}l@{\hspace{6pt}}
					@{\hspace{3pt}}r@{\hspace{6pt}}
					@{\hspace{3pt}}r@{\hspace{6pt}}
					@{\hspace{3pt}}r@{\hspace{6pt}}
					@{\hspace{3pt}}r@{\hspace{6pt}}
					@{\hspace{3pt}}r@{\hspace{6pt}}
					@{\hspace{3pt}}r@{\hspace{6pt}}
					@{\hspace{3pt}}r@{\hspace{6pt}}
				}
				\toprule
				Model & loss & {$\avgrank_{100}$}  & {$\hr_{100}$} & {$\avgrank_{500}$} & {$\hr_{500}$} 
				& {\auc} & {$\roc_5$} & {$\roc_{10}$} \\
				\midrule        
				\multirow{4}{*}{\convmodel}
				& \hingeone &   5.72 & 6.89 & 3.15 & 5.66 & 7.11 & 12.78 & 10.81\\
				& \hingetwo &   4.89 & 5.54 & 2.47 & 3.50 & 6.27 & 7.34 & 8.35\\
				& \hingethree & 4.75 & 5.88 & 0.93 & 3.62 & 6.24 & 7.21 & 7.29\\
				& \meansquare & -8.57 & 0.21 & -8.00 & -1.23 & 1.65 & -2.45 & -2.47\\
				
				%\cmidrule(lr){2-9}
				\midrule
				\multirow{4}{*}{\spannyconvmodel} 
				& \hingeone &   9.86 & \textbf{8.85} & \textbf{7.89} & 8.59 & \textbf{9.12} & \textbf{19.29} & \textbf{18.11}\\
				& \hingetwo &   8.18 & 7.64 & 6.46 & 6.64 & 8.44 & 15.59 & 14.08\\
				& \hingethree & 7.35 & 8.5 & 5.02 & 6.41 & 7.96 & 13.21 & 14.04\\
				& \meansquare & 7.10 & 8.05 & 2.49 & 4.95 & 6.30 & 15.17 & 12.5\\
				%	\cmidrule(lr){2-9}
				\midrule
				\multirow{4}{*}{\mhcflurry }
				& \hingeone &   \textbf{10.11} & 8.54 & 7.14 & \textbf{9.21} & 6.09 & 16.00 & 14.65\\
				& \hingetwo &   8.78 & 6.60 & 5.55 & 5.95 & 5.50 & 13.19 & 11.30\\
				& \hingethree & 6.50 & 6.20 & 5.08 & 5.47 & 4.69 & 12.19 & 8.12\\
				& \meansquare & 0.00 & 0.00 & 0.00 & 0.00 & 0.00 & 0.00 & 0.00\\
				\bottomrule
			\end{tabular}
			\begin{footnotesize}
				\begin{tablenotes}
					\item 
					\!\!The values in the table are percentage improvement compared with the baseline  \mhcflurry with \meansquare. 
					Models are trained using  $\blosum$+$\onehot$+$\deep$ encoding methods, and 
					selected with respect to $\hr_{100}$ and evaluated using the 7 evaluation metrics. 
					The best improvement with respect to each metric is \textbf{bold}. 
				\end{tablenotes}
			\end{footnotesize}
		\end{threeparttable}
	\end{small}  
\end{table}
%\label{tbl:perform:hr100} 
%
%
%
%\input{tables/blosum_onehot_learned_aucresults}
\begin{table}[!h]
	%  \vspace{-5pt}    
	\centering
	\caption{Performance Comparison over \mhcflurry with \meansquare  ($\auc$; $\blosum$+$\onehot$+$\deep$)}
	\label{tbl:perform:auc}
	%\vspace{-10pt}
	%\resizebox{\columnwidth}{!}{
	\begin{small}
		\begin{threeparttable}
			\begin{tabular}{
					@{\hspace{3pt}}l@{\hspace{6pt}}
					@{\hspace{3pt}}l@{\hspace{6pt}}
					@{\hspace{3pt}}r@{\hspace{6pt}}
					@{\hspace{3pt}}r@{\hspace{6pt}}
					@{\hspace{3pt}}r@{\hspace{6pt}}
					@{\hspace{3pt}}r@{\hspace{6pt}}
					@{\hspace{3pt}}r@{\hspace{6pt}}
					@{\hspace{3pt}}r@{\hspace{6pt}}
					@{\hspace{3pt}}r@{\hspace{6pt}}
				}
				\toprule
				Model & loss & {$\avgrank_{100}$}  & {$\hr_{100}$} & {$\avgrank_{500}$} & {$\hr_{500}$} 
				& {\auc} & {$\roc_5$} & {$\roc_{10}$} \\
				\midrule        
				\multirow{4}{*}{\convmodel}
				& \hingeone &   6.83 & 2.28 & 5.03 & 7.34 & 3.44 & 9.69 & 8.55\\
				& \hingetwo &   3.00 & 3.12 & 3.88 & 6.88 & 3.16 & 5.41 & 5.81\\
				& \hingethree & 3.66 & -0.63 & 3.01 & 5.83 & 2.81 & 4.43 & 3.58\\
				& \meansquare & -9.12 & -8.45 & -5.46 & -2.31 & -0.52 & -5.62 & -5.45\\
				
				%\cmidrule(lr){2-9}
				\midrule
				\multirow{4}{*}{\spannyconvmodel} 
				& \hingeone &   \textbf{10.27} & \textbf{4.42} & \textbf{9.24} & \textbf{10.08} & \textbf{5.09} & \textbf{17.20} & \textbf{15.06}\\
				& \hingetwo &   7.18 & 2.84 & 8.12 & 8.31 & 4.42 & 12.36 & 10.63\\
				& \hingethree & 7.54 & 2.98 & 7.69 & 7.25 & 4.44 & 11.57 & 10.04\\
				& \meansquare & 4.31 & 2.49 & 5.28 & 4.78 & 2.86 & 11.68 & 9.47\\
				%	\cmidrule(lr){2-9}
				\midrule
				\multirow{4}{*}{\mhcflurry }
				& \hingeone &   8.08 & 1.61 & 7.45 & 8.35 & 4.3 & 11.02 & 11.05\\
				& \hingetwo &   7.64 & 3.96 & 6.31 & 7.61 & 3.56 & 8.63 & 7.63\\
				& \hingethree & 7.39 & 1.64 & 5.81 & 6.74 & 3.41 & 9.32 & 7.68\\
				& \meansquare & 0.00 & 0.00 & 0.00 & 0.00 & 0.00 & 0.00 & 0.00\\
				\bottomrule
			\end{tabular}
			\begin{footnotesize}
				\begin{tablenotes}
					\item 
					\!\!The values in the table are percentage improvement compared with the baseline  \mhcflurry with \meansquare. 
					Models are trained using  $\blosum$+$\onehot$+$\deep$ encoding methods, and 
					selected with respect to $\auc$ and evaluated using the 7 evaluation metrics. 
					The best improvement with respect to each metric is \textbf{bold}. 
				\end{tablenotes}
			\end{footnotesize}
		\end{threeparttable}
		%}
	\end{small}
	%  \vspace{-10pt}    
\end{table}
%\label{tbl:perform:auc} 
%
%
%
\begin{table}[!h]
	%  \vspace{-5pt}    
	\centering
	\caption{Performance Comparison over \mhcflurry with \meansquare  ($\roc_{5}$; $\blosum$+$\onehot$+$\deep$)}
	\label{tbl:perform:roc}
	%\vspace{-10pt}
	\begin{small}
		\begin{threeparttable}
			\begin{tabular}{
					@{\hspace{3pt}}l@{\hspace{6pt}}
					@{\hspace{3pt}}l@{\hspace{6pt}}
					@{\hspace{3pt}}r@{\hspace{6pt}}
					@{\hspace{3pt}}r@{\hspace{6pt}}
					@{\hspace{3pt}}r@{\hspace{6pt}}
					@{\hspace{3pt}}r@{\hspace{6pt}}
					@{\hspace{3pt}}r@{\hspace{6pt}}
					@{\hspace{3pt}}r@{\hspace{6pt}}
					@{\hspace{3pt}}r@{\hspace{6pt}}	
				}
				\toprule
				Model & loss & {$\avgrank_{100}$}  & {$\hr_{100}$} & {$\avgrank_{500}$} & {$\hr_{500}$} 
				& {\auc} & {$\roc_5$} & {$\roc_{10}$} \\
				\midrule        
				\multirow{4}{*}{\convmodel}
				& \hingeone & 8.38 & 6.71 & 3.23 & 6.86 & 5.14 & 5.91 & 4.63 \\ 
				& \hingetwo & 3.33 & 5.16 & 0.96 & 5.52 & 4.22 & 5.22 & 2.56 \\ 
				& \hingethree & 8.36 & 7.95 & 3.35 & 5.83 & 5.01 & 5.40 & 3.70 \\ 
				& \meansquare & -11.59 & -5.75 & -8.19 & -2.14 & 0.23 & -7.20 & -8.84 \\ 
				
				%\cmidrule(lr){2-9}
				\midrule
				\multirow{4}{*}{\spannyconvmodel} 
				& \hingeone & \textbf{12.78} & \textbf{15.81} & \textbf{8.31} & \textbf{11.11} & \textbf{7.29} & \textbf{15.03} & \textbf{10.52} \\ 
				& \hingetwo & 11.71 & 9.88 & 5.45 & 8.12 & 6.20 & 12.09 & 7.72 \\ 
				& \hingethree & 10.18 & 8.09 & 7.40 & 9.58 & 6.65 & 10.38 & 8.09 \\ 
				& \meansquare & 5.25 & 3.68 & 2.96 & 4.81 & 4.49 & 13.54 & 6.87 \\ 
				%	\cmidrule(lr){2-9}
				\midrule
				\multirow{4}{*}{\mhcflurry }
				& \hingeone & 11.20 & 11.53 & 7.09 & 7.44 & 6.32 & 10.38 & 8.31 \\ 
				& \hingetwo & 1.87 & 0.36 & 0.09 & 1.16 & 3.22 & 13.76 & 9.80 \\ 
				& \hingethree & 7.64 & 8.25 & 5.83 & 6.28 & 5.36 & 8.66 & 7.74 \\ 
				& \meansquare & 0.00 & 0.00 & 0.00 & 0.00 & 0.00 & 0.00 & 0.00 \\
				\bottomrule
			\end{tabular}
			\begin{footnotesize}
				\begin{tablenotes}
					\item 
					\!\!The values in the table are percentage improvement compared with the baseline  \mhcflurry with \meansquare. 
					Models are trained using  $\blosum$+$\onehot$+$\deep$ encoding methods, and 
					selected with respect to $\roc_{5}$ and evaluated using the 7 evaluation metrics. 
					The best improvement with respect to each metric is \textbf{bold}. 
					\par
				\end{tablenotes}
			\end{footnotesize}
		\end{threeparttable}
	\end{small}
	%  \vspace{-10pt}    
\end{table}
%label{tbl:perform:roc}
%
%
%
%\input{spanny_embed_avg100results}
\begin{table}[!h]
	%  \vspace{-5pt}    
	\centering
	\caption{Encoding Performance Comparison on \spannyconvmodel with \hingeone ($\avgrank_{100}$)}
	\label{tbl:perform:embed_ar100}
	%\vspace{-10pt}
	\begin{small}
		\begin{threeparttable}
			\begin{tabular}{
					@{\hspace{3pt}}l@{\hspace{6pt}}
					@{\hspace{3pt}}r@{\hspace{6pt}}
					@{\hspace{3pt}}r@{\hspace{6pt}}
					@{\hspace{3pt}}r@{\hspace{6pt}}
					@{\hspace{3pt}}r@{\hspace{6pt}}
					@{\hspace{3pt}}r@{\hspace{6pt}}
					@{\hspace{3pt}}r@{\hspace{6pt}}
					@{\hspace{3pt}}r@{\hspace{6pt}}
					@{\hspace{3pt}}r@{\hspace{6pt}}
				}
				\toprule
				encoding & {$\avgrank_{100}$}  & {$\hr_{100}$} & {$\avgrank_{500}$} & {$\hr_{500}$} 
				%encoding & \footnotesize{AR100}  & \footnotesize{HR100} & \footnotesize{AR500} & \footnotesize{HR500} 
				%& \footnotesize{AUC} & \footnotesize{ROC5} & \footnotesize{ROC10} \\
				& {\auc} & {$\roc_5$} & {$\roc_{10}$} \\  
				
				\midrule        
				%\multirow{4}{*}{\convmodel}
				\blosum &       0.00 & 0.00 & 0.00 & 0.00 & 0.00 & 0.00 & 0.00\\
				\onehot &       -6.04 & -4.09 & -6.00 & -1.75 & -1.13 & -5.38 & -5.34\\
				\deep &         -4.74 & -0.98 & -2.55 & -0.09 & -0.13 & 0.26 & -1.17\\
				\blosum+\onehot &       -0.62 & -0.06 & -0.62 & 1.68 & 0.27 & 3.17 & 1.22\\
				\blosum+\deep & \textbf{1.43} & \textbf{3.33} & \textbf{0.66} & 2.62 & \textbf{0.57} & 4.88 & 2.73\\
				\onehot+\deep & -4.10 & -3.16 & -2.64 & -0.37 & 0.01 & -1.83 & -2.29\\
				\blosum+\onehot+\deep & 0.15 & -0.15 & -0.47 & \textbf{3.53} & 0.53 & \textbf{6.29} & \textbf{3.53}\\
				\bottomrule
			\end{tabular}
			\begin{footnotesize}
				\begin{tablenotes}
					\item 
					\!\!The values in the table are percentage improvement compared with the \blosum encoding method. 
					Models are selected with respect to $\avgrank_{100}$ and evaluated using the 7 evaluation metrics. 
					The best improvement with respect to each metric is \textbf{bold}. 
					\par
				\end{tablenotes}
			\end{footnotesize}
		\end{threeparttable}
	\end{small}
	%  \vspace{-10pt}    
\end{table}
%\label{tbl:perform:embed_ar100}
%
%
%
%\input{spanny_embed_hr100results}
\begin{table}[!h]
	%  \vspace{-5pt}    
	\centering
	\caption{Encoding Performance Comparison on \spannyconvmodel with \hingeone ($\hr_{100}$)}
	\label{tbl:perform:embed_hr100}
	%\vspace{-10pt}
	\begin{small}
		\begin{threeparttable}
			\begin{tabular}{
					@{\hspace{3pt}}l@{\hspace{6pt}}
					@{\hspace{3pt}}r@{\hspace{6pt}}
					@{\hspace{3pt}}r@{\hspace{6pt}}
					@{\hspace{3pt}}r@{\hspace{6pt}}
					@{\hspace{3pt}}r@{\hspace{6pt}}
					@{\hspace{3pt}}r@{\hspace{6pt}}
					@{\hspace{3pt}}r@{\hspace{6pt}}
					@{\hspace{3pt}}r@{\hspace{6pt}}
					@{\hspace{3pt}}r@{\hspace{6pt}}
				}
				\toprule
				encoding & {$\avgrank_{100}$}  & {$\hr_{100}$} & {$\avgrank_{500}$} & {$\hr_{500}$} 
				%encoding & \footnotesize{AR100}  & \footnotesize{HR100} & \footnotesize{AR500} & \footnotesize{HR500} 
				%& \footnotesize{AUC} & \footnotesize{ROC5} & \footnotesize{ROC10} \\
				& {\auc} & {$\roc_5$} & {$\roc_{10}$} \\  
				
				\midrule        
				%\multirow{4}{*}{\convmodel}
				\blosum &       0.00 & 0.00 & 0.00 & 0.00 & 0.00 & 0.00 & 0.00\\
				\onehot &       -6.20 & -3.31 & -5.71 & -3.85 & -0.69 & -7.10 & -5.47\\
				\deep &         -3.81 & -0.44 & -2.64 & -0.21 & 0.26 & -0.36 & 0.09\\
				\blosum+\onehot &       -1.47 & 0.98 & -0.70 & -0.04 & 0.11 & 1.61 & 0.85\\
				\blosum+\deep & -0.40 & \textbf{2.45} & -0.30 & 0.57 & 0.28 & 3.39 & 1.97\\
				\onehot+\deep & -6.45 & -2.48 & -4.34 & -1.71 & -0.55 & -3.08 & -3.13\\
				\blosum+\onehot+\deep & \textbf{0.58} & 0.07 & \textbf{0.98} & \textbf{2.16} & \textbf{0.75} & \textbf{4.56} & \textbf{3.50}\\
				\bottomrule
			\end{tabular}
			\begin{footnotesize}
				\begin{tablenotes}
					\item 
					\!\!The values in the table are percentage improvement compared with the \blosum encoding method. 
					Models are selected with respect to $\hr_{100}$ and evaluated using the 7 evaluation metrics. 
					The best improvement with respect to each metric is \textbf{bold}. 
					\par
				\end{tablenotes}
			\end{footnotesize}
		\end{threeparttable}
	\end{small}
	%  \vspace{-10pt}    
\end{table}
%\label{tbl:perform:embed_roc5}
%
%
%
%\input{spanny_embed_aucresults}
\begin{table}[!h]
	%  \vspace{-5pt}    
	\centering
	\caption{Encoding Performance Comparison on \spannyconvmodel with \hingeone (\auc)}
	\label{tbl:perform:encode_auc}
	%\vspace{-10pt}
	\begin{small}
		\begin{threeparttable}
			\begin{tabular}{
					@{\hspace{3pt}}l@{\hspace{6pt}}
					@{\hspace{3pt}}r@{\hspace{6pt}}
					@{\hspace{3pt}}r@{\hspace{6pt}}
					@{\hspace{3pt}}r@{\hspace{6pt}}
					@{\hspace{3pt}}r@{\hspace{6pt}}
					@{\hspace{3pt}}r@{\hspace{6pt}}
					@{\hspace{3pt}}r@{\hspace{6pt}}
					@{\hspace{3pt}}r@{\hspace{6pt}}
					@{\hspace{3pt}}r@{\hspace{6pt}}
				}
				\toprule
				encoding & {$\avgrank_{100}$}  & {$\hr_{100}$} & {$\avgrank_{500}$} & {$\hr_{500}$} 
				%encoding & \footnotesize{AR100}  & \footnotesize{HR100} & \footnotesize{AR500} & \footnotesize{HR500} 
				%& \footnotesize{AUC} & \footnotesize{ROC5} & \footnotesize{ROC10} \\
				& {\auc} & {$\roc_5$} & {$\roc_{10}$} \\  
				
				\midrule        
				%\multirow{4}{*}{\convmodel}
				\blosum &       0.00 & 0.00 & 0.00 & 0.00 & 0.00 & 0.00 & 0.00\\
				\onehot &       -6.75 & -4.84 & -4.37 & -2.55 & -0.91 & -6.85 & -5.34\\
				\deep &         -3.31 & -0.59 & -2.09 & -0.71 & -0.19 & -1.06 & -1.49\\
				\blosum+\onehot &       -0.45 & \textbf{0.97} & 0.22 & 0.52 & 0.09 & 1.41 & 1.02\\
				\blosum+\deep & \textbf{0.95} & -0.84 & 1.09 & \textbf{1.58} & 0.55 & 3.25 & 2.00\\
				\onehot+\deep & -2.90 & -3.40 & -2.68 & -1.56 & -0.39 & -4.13 & -3.26\\
				\blosum+\onehot+\deep & 0.77 & -0.67 & \textbf{1.43} & 1.05 & \textbf{0.62} & \textbf{5.90} & \textbf{3.95}\\
				\bottomrule
			\end{tabular}
			\begin{footnotesize}
				\begin{tablenotes}
					\item 
					\!\!The values in the table are percentage improvement compared with the \blosum encoding method. 
					Models are selected with respect to $\auc$ and evaluated using the 7 evaluation metrics. 
					The best improvement with respect to each metric is \textbf{bold}. 
					\par
				\end{tablenotes}
			\end{footnotesize}
		\end{threeparttable}
	\end{small}
	%  \vspace{-10pt}    
\end{table}
%\label{tbl:perform:encode_auc}
%\input{spanny_embed_roc5}
\begin{table}[!h]
	%  \vspace{-5pt}    
	\centering
	\caption{Encoding Performance Comparison on \spannyconvmodel with \hingeone ($\roc_{5}$)}
	\label{tbl:perform:embed_roc5}
	\begin{small}
		\begin{threeparttable}
			\begin{tabular}{
					@{\hspace{3pt}}l@{\hspace{6pt}}
					@{\hspace{3pt}}r@{\hspace{6pt}}
					@{\hspace{3pt}}r@{\hspace{6pt}}
					@{\hspace{3pt}}r@{\hspace{6pt}}
					@{\hspace{3pt}}r@{\hspace{6pt}}
					@{\hspace{3pt}}r@{\hspace{6pt}}
					@{\hspace{3pt}}r@{\hspace{6pt}}
					@{\hspace{3pt}}r@{\hspace{6pt}}
					@{\hspace{3pt}}r@{\hspace{6pt}}
				}
				\toprule
				encoding & {$\avgrank_{100}$}  & {$\hr_{100}$} & {$\avgrank_{500}$} & {$\hr_{500}$} 
				%& \footnotesize{AUC} & \footnotesize{ROC5} & \footnotesize{ROC10} \\
				& {\auc} & {$\roc_5$} & {$\roc_{10}$} \\  
				
				\midrule        
				%\multirow{4}{*}{\convmodel}
				\blosum &       0.00 & 0.00 & 0.00 & 0.00 & 0.00 & 0.00 & 0.00\\
				\onehot &       -7.96 & -7.56 & -5.22 & -2.32 & -1.05 & -5.61 & -4.81\\
				\deep &         -2.29 & -0.84 & -2.76 & -0.78 & -0.54 & -0.69 & -1.28\\
				\blosum+\onehot &       1.24 & -0.16 & 0.18 & 0.40 & 0.00 & 1.38 & 1.12\\
				\blosum+\deep & \textbf{2.84} & 0.15 & 1.41 & \textbf{1.90} & 0.60 & 3.20 & 3.14\\
				\onehot+\deep & -4.76 & -2.90 & -3.31 & -1.28 & -0.61 & -2.58 & -2.30\\
				\blosum+\onehot+\deep & 2.57 & \textbf{0.20} & \textbf{2.08} & 1.46 & \textbf{0.82} & \textbf{3.83} & \textbf{3.47}\\
				\bottomrule
			\end{tabular}
			\begin{footnotesize}
				\begin{tablenotes}
					\item 
					\!\!The values in the table are percentage improvement compared with the \blosum encoding method. 
					Models are selected with respect to $\roc_{5}$ and evaluated using the 7 evaluation metrics. 
					The best improvement with respect to each metric is \textbf{bold}. 
					\par
				\end{tablenotes}
			\end{footnotesize}
		\end{threeparttable}
	\end{small} 
\end{table}
%\label{tbl:perform:embed_hr100}

\end{appendix}
%
%\clearpage
%\beginsupplement
%
%\section*{Supplemental Materials}
%\textcolor{blue}{I haven't revised this part yet...}
%\subsection*{Training Details}
%\textcolor{blue}{These hyper-parameters for {\convmodel} and {\spannyconvmodel} included the number of hidden units 
%	(16, 64, 128) in fully connected layer, the number of filters (16, 64, 128) and the kernel size 
%	(1, 3, 5) in convolutional layer (and global kernel) and the batch size ( 32, 64, 128).
%	%
%	The hyper-parameters for {\mhcflurry} included the number of hidden units (16, 64, 128) in fully 
%	connected layer, the number of filters (8, 64), the kernel size (3, 5) and the batch+ size (32, 64, 128).}
%\textcolor{blue}{The initial learning rate is set to be equal to 0.1.
%	%
%	We use ReduceLROnPlateau to reduce the learning rate by a factor $0.1$ if the loss on validation 
%	set hasn't decreased for 5 epochs. 
%	%
%	If the loss on validation set hasn't decreased for 20 epochs, the training process would be interrupted 
%	and early stopped.}
\end{document}